\documentclass[11pt,superscriptaddress,aps,prd,preprint]{revtex4}

\usepackage{amsfonts}
\usepackage{graphicx}
\usepackage{amssymb}
\usepackage{amsmath}
\usepackage{indentfirst}
\usepackage{graphicx}
\usepackage{subfigure}
\usepackage{float}
\usepackage{color}
\usepackage{amsmath,bm}

\begin{document}


\title{Non-abelian gauge symmetry for fields in phase space: a realization of the Seiberg-Witten non-abelian gauge theory }

\author{J. S. Cruz-Filho}
\affiliation{Secretaria de Estado de Educa\c{c}\~ao de Mato Grosso, 78049-909, Cuiab\'{a}, Mato Grosso, Brazil}
\author{R.G.G. Amorim}
\affiliation{International Center of Physics, Instituto de F\'{\i}sica, Universidade de
Bras\'{\i}lia, 70.910-900, Brasilia, DF, Brazil}
\affiliation{Faculdade Gama, Universidade de Bras\'{\i}lia, 72444-240, Brasilia, DF,
Brazil}
\author{F. C. Khanna}
\thanks{Present address: Department of Physics and Astronomy, University of
Victoria, BC V8P 5C2, Canada}
\affiliation{Physics Department, Theoretical Physics Institute, University of Alberta,
Edmonton, Alberta T6G 2J1 Canada}
\affiliation{TRIUMF, 4004, Westbrook Mall, Vancouver, British Columbia V6T 2A3, Canada}
\author{A. E. Santana}
\affiliation{International Center of Physics, Instituto de F\'{\i}sica, Universidade de
Bras\'{\i}lia, 70.910-900, Brasilia, DF, Brazil}
\author{A. F. Santos}
\affiliation{Instituto de F\'{\i}sica, Universidade Federal de Mato Grosso,\\
78060-900, Cuiab\'{a}, Mato Grosso, Brazil}
\author{S. C. Ulhoa}
\affiliation{International Center of Physics, Instituto de F\'{\i}sica, Universidade de
Bras\'{\i}lia, 70.910-900, Brasilia, DF, Brazil}

\begin{abstract}
The Seiberg-Witten formalism has been realized as an electrodynamics in phase space (associated to the Dirac equation written in phase space) and this fact is explored here with non-abelian gauge group. First,  a physically heuristic presentation of the Seiberg-Witten approach is carried out for non-abelian gauge in order to guide the calculation procedures. These results are realized by starting with the  Lagrangian density for the free Dirac field in phase space. Then a  field strength  is derived, where the non-abelian gauge group is the SU(2), corresponding to an isospin (non-abelian) field theory in phase space. An application to nucleon is then discussed.

\end{abstract}

\maketitle

\date{Jan. 28, 2018}


\section{Introduction}

The development of quantum mechanics from classical mechanics initiated with the experimental observation that
position and momentum coordinates do not commute. This had a tremendous impact
on the development of dynamics of a broad variety of areas, including quantum corrections to statistical mechanics in phase space, a program that  had its origin with Wigner~\cite{Wigner}.
In addition, there was the question about non-commutation of position coordinates, a conjecture raised by
Heisenberg~\cite{Heisenberg}, and first developed by Snyder~\cite{Snyder},  studying representations of Lie-groups. This was followed by advances in non-commutative geometry, having as a realization the algebraic structure of the Wigner approach, based on the non-commutative Moyal (star)-product in phase space~\cite{Moyal,Hillery,zabo1,Grosse}.

Then the problem of a non-commutative quantum field came about. In this case,  the abelian gauge symmetry for non-commutative fields was addressed consistently first by Seiberg and Witten~\cite{SeiberWitten}, that used this for string theory. The usual gauge-field strength is then generalized to a spin-one non-commutative field theory and the quantum field approach gives rise to a mixture of infrared and ultraviolet divergences that usually
breakdown renormalizability~\cite{zabo1,Minwalla}.

The difficulties with the  gauge non-commutative fields have been handled~\cite{Langmann, Magnen}, and numerous applications of non-commutative theories and its algebraic structure have been achieved, considering, for instance: the field theory of the standard model of particle physics, including the development of the non-commutative non-abelian theory~\cite{zabo1, Mariz, Costa, Gurau}; the
gravitational theory~\cite{Kalau, Kastler}; supersymmetry~\cite{Girotti}; quantum Hall effect~\cite{Belissard};
unification of gravity and quantum field theory~\cite{Yu}; and the pair-creation of neutral Dirac particles in
$(1+2)d$ non-commutative space-time have been investigated~\cite{Hammil}.

A realization of the Seiberg-Witten-gauge theory has been derived by  exploring a U(1)-unitary representation of the Poincar\'e space-time symmetry in  phase space~\cite{Amorim0}, giving rise to a symplectic quantum electrodynamics  where the propagators are associated with the Wigner function~\cite{Amorim19}. The non-commutativity, in this case, is among the two sectors of the symplectic manifold; that is, among coordinates and momenta. In such a symplectic realization, the  field strength is given by the algebraic non-commutative structure of the  Moyal (star)-product~\cite{Moyal,Hillery}. Other representations, such as spin zero and spin 1/2, have been explored for the Galilei and Poincar\'e groups~\cite{Oliveira, Ronni}, leading to the Schr\"{o}dinger equation in phase space, in the non-relativistic realm,  as well as for the  Klein-Gordon and Dirac equation in a relativistic phase space. The Wigner function is given by $f_{w}(q,p)=~\psi (q,p) \star  \psi ^{\dag }(q,p)$, where $\psi (q,p) $
stands for the wave function in phase space called the quasi-amplitudes of probability and ``$\star $" is the Moyal product in this phase space.

One basic motivation for such developments is the calculation of the Wigner function for the states of relativistic particles described by a quantum field~\cite{Amorim0,Amorim19}. This is  important for analysing the statistical nature (such as chaoticity) of quantum states, with plain  measurement in some systems~\cite{wig3,wig4,2gal1,2davido1,2kha1}. Despite the importance, the direct application of the Wigner formalism is not a straightforward procedure. For instance, considering interaction, neither the Liouville-von Neumann equation  is trivially generalized for a quantum field, nor the gauge symmetry is a simple task to be taken into consideration, since the Wigner function is a Real function.  These problems  have led to looking for solution in a representation formalism based on unitary transformations in phase space, such that the state, the quasi-amplitude of probability,  is consistently associated with the Wigner function. This has been accomplished~\cite{Oliveira, Ronni}, and in this approach, the Seiberg-Witten-gauge theory arises  naturally by studying gauge transformations of the type $\psi(q,p)\rightarrow e^{-i\lambda(q,p)} \star \psi(q,p)$, leading to a non-commutative-like electrodynamics, such that the propagators, describing bosons and fermions, are related to the Wigner function~\cite{Amorim0}. Applications of such results include:   the interaction between the Dirac equation with an external electromagnetic field in phase space~\cite{Ronni001}; analysis of the Wigner function for the Landau problem~\cite{Ronni002}; evaluation of the negativity of the Wigner function for a system defined by the sum of H\'enon-Heiles potential and Hydrogen atom~\cite{Cruz}.

Despite these advances with the field theory in phase phase, the non-abelian gauge symmetry has not been developed for fields in phase space. This problem is addressed here, first by following a heuristic presentation of the Seiberg-Witten approach, to include a non-abelian gauge group in the non-commutative  field strength in phase space. This analysis is  carried out in Section 2, where the notation is fixed and some aspects of a non-commutative field are outlined. In Section 3, starting with the  Lagrangian density for the free Dirac field in phase space, and considering the SU(2) as the non-abelian gauge group, a field strength is derived, corresponding to a realization of the Lagrangian and the field strength discussed in Section 2.  In Section 4, an application to the nucleon in an external field is presented. Final concluding remarks are presented in Section 5. In  Appendix A, some aspects of the symplectic field theory are outlined in order to fix the notation. In Appendix B, some details of the calculation for the abelian symplectic gauge field is presented for the sake of completeness and for showing, as a guide, some steps and properties used in Section 3.

\section{Gauge theory for non-commutative fields}

The non-commutative plane or the Groenewold-Moyal (GM) plane is defined by
the following procedure. Consider the manifold $\mathbb{R}^{d}$ with
commutative coordinates $x_{\mu }$,  $\mu =1, ..., d$,  i.e. $\left[ x^{\mu
},x^{\nu }\right] =0.$ The space of complex functions is then introduced: $%
\mathcal{A}_{0}(\mathbb{R}^{d})=\{f:\mathbb{R}^{d}\rightarrow \mathbb{C}\}$,
such that $\mathcal{A}_{0}(\mathbb{R}^{d})$ is an associative algebra with
multiplication $f\cdot g(x):=f(x)g(x)$. There is then a 1:1 map between the
manifold $\mathbb{R}^{d}$ and the algebra $\mathcal{A}_{0}(\mathbb{R}^{d})$.

A deformed algebra of operators $\mathbb{R}_{\theta }^{d}$ is generated by
the  $d$ operators $\hat{x}^{\mu }$ with the deformed relations
\begin{equation}
\left[ \hat{x}^{\mu },\hat{x}^{\nu }\right] =i\theta ^{\mu \nu },
\end{equation}%
with $\theta ^{\mu \nu }=-\theta ^{\nu \mu }\ $\ being real constants. The Weyl quantization is a mapping from
elements of the complex vector space $\mathcal{A}_{0}(\mathbb{R}^{d})$ to
elements of the algebra $\mathbb{R}_{\theta }^{d}$ through the Weyl map $%
\hat{W}$, which is defined by $\hat{W}:\mathcal{A}_{0}(\mathbb{R}%
^{d})\rightarrow \mathbb{R}_{\theta }^{d}$, such that $e_{k}(x)=\exp
(ik_{\mu }x^{\mu })\mapsto e_{k}(\hat{x})\equiv \exp (ik_{\mu }\hat{x}^{\mu
})$ and with
\begin{equation}
\hat{f}\equiv \hat{W}[f]=\int \frac{d^{d}k}{(2\pi )^{d}}~\tilde{f}(k)e_{k}(%
\hat{x}),
\end{equation}%
where $~f(k)$ is the Fourier transform of $~f(x)$; i.e$.$%
\begin{equation}
f(x)=\int \frac{d^{d}k}{(2\pi )^{d}}~f(k)e_{k}(x).
\end{equation}%
The operator $\hat{W}[f]$ is a non-commutative field.

The inverse of the Weyl map, $W,$ is called the Wigner function, which  is
defined by $\hat{W}^{-1}:\mathbb{R}_{\theta }^{d}\rightarrow \mathcal{A}_{0}(%
\mathbb{R}^{d})$; or $\hat{f}\mapsto \hat{W}^{-1}[\hat{f}]$, such that
\begin{equation}
f(x)\equiv \hat{W}^{-1}[\hat{f}]=\int \frac{d^{d}k}{(2\pi )^{d}}e^{-ik_{\mu
}x^{\mu }}\mathrm{Tr}(\hat{f}~e_{k}(\hat{x})),
\end{equation}%
where $\mathrm{Tr}$ is the trace. The  map $W$  provides a vector space (not
an algebra) isomorphism. Then an algebra $\mathcal{A}_{\theta }(\mathbb{R}%
^{d})$ is introduced with  the Moyal product or $\star $-product, given by
\begin{equation}
f\star g=\hat{W}^{-1}[\hat{W}[f]\hat{W}[g]],\ \ f,g\in \mathcal{A}_{\theta }(%
\mathbb{R}^{d}),
\end{equation}%
such that for $\theta ^{\mu \nu }\mapsto 0,$ the product $(f\star
g)(x)=f(x)\star g(x)\mapsto f(x)g(x)$. This Moyal product is associative,
but it is not commutative. Then $\mathcal{A}_{\theta }(\mathbb{R}^{d})$ is a
non-commutative algebra defined from the  commutative plane $\mathcal{A}_{0}(%
\mathbb{R}^{d}).$  Then the algebra $\mathcal{A}_{\theta }(\mathbb{R}^{d})$
is called a non-commutative plane or the Groenewold-Moyal (GM) plane. As an
example, for $\ C^{\infty }$ functions in $\mathcal{A}_{\theta }(\mathbb{R}%
^{d})$, a realization of the  Moyal product is given by
\begin{equation}
(f\star g)(x)=f(x)e^{\frac{i}{2}\overleftarrow{\partial }_{\mu }\theta ^{\mu
\nu }\overrightarrow{\partial }_{\nu }}g(x).
\end{equation}

This structure is the starting point for developing the non-commutative
field theory; and considering the association of non-commutative geometry
with string theory, Seiberg and Witten~\cite{SeiberWitten} studied a string
dynamics described by a minimally coupled supersymmetric gauge field in a
non-commutative space. The result is a generalized gauge field $A^{\mu }$
with an antisymmetric field strength  given by
\begin{equation}
F_{\theta }^{\mu \nu }=\partial _{\nu }A^{\mu }-\partial _{\mu }A^{\nu
}-i\{A^{\mu },A^{\nu }\}_{M},  \label{sw1}
\end{equation}%
where $\{A^{\mu },A^{\nu }\}_{M}$ is the Moyal bracket, given by $%
\{f,g\}_{M}=f\star g-g\star f$. It is to be noted that $\theta ^{\mu \nu }\mapsto
0,\ \ \{f,g\}_{M}\mapsto 0,\ $ i.e., $F_{\theta }^{\mu \nu }\mapsto F^{\mu
\nu }=\partial _{\nu }A^{\mu }-\partial _{\mu }A^{\nu }$; that is, the usual
electrodynamics (the U$(1)$ gauge-group theory) is recovered. In this case, the starting point is the (3+1)-Minkowski space, $\mathbb M$, with $\mu,\nu=0,1,2,3$, and the metric $g$ being such that $diag (g) = (1,-1,-1,-1)$.  This
analysis  provides the hint for writing  the Seiberg-Witten theory considering
non-abelian fields, with the non-abelian group characterized by the
structure constants $c_{rsl}$~\cite{zabo1}.

 The  phase transformation of a non-commutative field can be written as $%
\psi ^{\prime }(x)=U(x;\lambda ) \star \psi (x),$ where $  U(x;\lambda )=e^{i\lambda(x) }$,
with $\lambda (x) $  being a real-valued operator of the space-time
coordinates. The transformation $U(x;\lambda )$ is unitary. Considering components of $\psi _{j}(x)$,  $j=1,2,...,N$ , the transformation $U$ is an
$N\times N$ matrix in the indices $i$ and $j$, such that
\begin{equation}
\psi _{i}^{\prime }(x)=U_{ij}(x;\lambda )\star \psi _{j}(x)\ \ \ \ \text{and}%
\ \ \ \overline{\psi }_{i}^{\prime }(x)=\overline{\psi }_{j}(x)\star U_{ji}^{\dagger
}(x;\lambda ),  \label{dir82}
\end{equation}%
where repeated Latin indices are summed. As $%
U_{ij}(x;\lambda )$ is connected to identity, we write
$
U(x;\lambda )=e^{-i\xi\lambda (x)}=e^{-i\xi\Lambda _{r}(x)t_{r}},
$
where  $\lambda_{r}(x)$ are real-valued functions of space-time
coordinates and $\xi$ is a constant fixing the units. The quantities  $t_{j}$
are operators of the gauge group, satisfying the Lie algebra $%
[t_{r},t_{s}]=c_{rsl}t_{l}$, where $c_{rsl}$ are the structure constants of
the group, with $r,s,l=1,2,...,N$, and $N$ stands for the number of
group generators $t_{r}$, taken in the adjoint representation. Then each
matrix $t_{r}$ is given in the form $(t_{r})_{sl}=(c_{rsl})$.

The non-abelian Seiberg-Witten gauge theory  is carried out by using specific
non-commutative field theories. It is expected that for $c_{rsl}=0$, the field strength
reduces to eq.~(\ref{sw1}). The natural candidate is a  field strength  given by
\begin{equation}
F_{\theta }^{\mu \nu }=\partial _{\nu }A^{\mu }-\partial _{\mu }A^{\nu
}-i[A^{\mu },A^{\nu }]_{M},
\end{equation}%
where  $[A^{\mu },A^{\nu }]_{M}=A^{\mu }\star A^{\nu }-A^{\nu }\star A^{\mu
}$ (it is important to emphasize that the fields $A^{\mu }$ are
matrices, as usual). Under a gauge transformation, the gauge field transforms as $A_{\mu
}^{^{\prime }}=A_{\mu }+[iA_{\mu },\lambda ]_{M}-\frac{i}{\xi}\partial _{\mu
}\lambda $. In addition, observe
that for  $c_{rsl}=0,$ {\ }$[A^{\mu },A^{\nu }]_{M}\mapsto \{f,g\}_{M}.$
Consistently, for non-zero structure constants, and for $\theta ^{\mu \nu
}\mapsto 0,\ \ \ [A^{\mu },A^{\nu }]_{M}\mapsto \lbrack A^{\mu },A^{\nu
}]=A^{\mu }A^{\nu }-A^{\nu }A^{\mu }.$  With this field strength, the simplest non-abelian  Lagrangian density reads
\begin{equation*}
\mathcal{L}=\frac{1}{4}F_{\theta }^{\mu \nu }F_{\theta\mu \nu }.
\end{equation*}
It has to be noted that there is no Moyal product in the product of two field strength, as usual in non-commutative theories~\cite{zabo1}.

In the next  section, these results are demonstrated explicitly for the
symplectic field theory, a non-commutative-like field theory in phase space,
associated with the Wigner function formalism. In this approach, the starting manifold is the cotangent-bundle of the Minkowski space, $T^{\ast }\mathbb{M}$,
where each point is specified by the coordinates
$(q^{\mu },p^{\mu }) \in T^{\ast }\mathbb{M} =\mathbb{M}\times \mathbb{M}$. The 8-dimensional space $T^{\ast }\mathbb{M}$ is
equipped with
 a 2-form $\Omega =dq^{\mu }\wedge dp_{\mu }$%
, called the symplectic form.  The space $T^{\ast }\mathbb{M}$
endowed with this symplectic structure is called the symplectic (or phase)
space, and will be denoted by $\Gamma $. Let us define a vector in this manifold by $%
\omega =({\omega }^{1},{\omega }^{2},\ldots ,{\omega }^{8})$, with $\omega
^{1}=q^{0},\omega ^{2}=q^{1},\omega ^{3}=q^{2},\omega ^{4}=q^{3},$ $\omega
^{5}=p^{0},\omega ^{6}=p^{1},\omega ^{7}=p^{2},\omega ^{8}=p^{3}$ with $%
q=(q^{0},\mathbf{q})$ and $p=(p^{0},\mathbf{p})$ being vectors in $\mathbb{M}
$. The symplectic metric matrix in $\Gamma $ is given by (${\eta }_{ab})$, $%
a,b=1,...,8$,
\begin{equation*}
\eta =\left(
\begin{array}{cc}
0 & I \\
-I & 0%
\end{array}%
\right) ,
\end{equation*}%
where $I$ is a $4\otimes4$ unit matrix. We define the following vector fields,
   operators on $C^{\infty
}(T^{\ast }M)$,
\begin{equation}
\Lambda =\eta_{ab}\frac{\overleftarrow{\partial }}{\partial \omega^{a }}\frac{%
\overrightarrow{\partial }}{\partial \omega^{b }},
\label{fasenova2}
\end{equation}%
such that for $C^{\infty }$ functions, $f(\omega)=f(q,p)$ and $g(\omega)=g(q,p),$ we have
\begin{eqnarray*}
\Omega (f\Lambda ,g\Lambda )&=&{\eta }^{ab}\frac{\partial {f}}{\partial {%
\omega ^{a}}}\frac{\partial {g}}{\partial {\omega ^{b}}}={\eta }%
^{ab}\partial _{a}{f}\partial _{b}{g}\\
&=&f\Lambda g=\{f,g\},
\end{eqnarray*}
where $\{f,g\}=\frac{\partial f}{\partial q^{\mu }}\frac{\partial g}{%
\partial p_{\mu }}-\frac{\partial f}{\partial p^{\mu }}\frac{\partial g}{%
\partial q_{\mu }}$ is the Poisson bracket.

In this structure, a Hilbert space is introduced  and used as a carrier space for the Poincar\'e symmetry. The basic unitary operators are given by
 \[F=f(\omega)\star =f(\omega)e^{%
\frac{i}{2}\overleftarrow{\partial }_{a }\hbar \eta ^{ab }\overrightarrow{%
\partial }_{b }},
\]
where  $\hbar$ is the Planck constant and the star-notation for the Moyal product is used as  $\star=\exp^{%
\frac{i}{2}\overleftarrow{\partial }_{a } \hbar \eta ^{ab }\overrightarrow{%
\partial }_{b }}$,
with $\theta^{ab}=\hbar \eta ^{ab }$.

Representations for the Poincar\'e symmetry are taken by considering the function $f$ to be components of  $q$ and $p$. Explicitly, we have
 \begin{eqnarray}
P^{\mu } &=&p_{\star}^{\mu }=p^{\mu }-\frac{%
i\hbar }{2}\frac{\partial }{\partial q_{\mu }}, \label{jan20143} \\
Q^{\mu } &=&q_{\star}^{\mu }=q^{\mu }+\frac{%
i \hbar}{2}\frac{\partial }{\partial p_{\mu }}.\label{jan20144}
\end{eqnarray}%
 These operators satisfy the Heinsenberg condition, i.e. $ [ Q^{\mu },P^{\nu }]=i \hbar g^{\mu \nu }$ .
Therefore,   Lorentz transformations are introduced by defining the generators
$
M_{\mu \nu }=Q^{\mu }P^{\nu }-P^{\nu }Q^{\mu }
$. It is simple to verify the physical consistency of the representation, $P$ is taken as the generator of translation in the Minkowski space. Then we have for $Q$ and $P$ the transformation rules of position and momentum, respectively. Indeed:
 $%
U(a)=\exp[-ia_{\mu }{P}^{\mu }]$ such that $U(a){Q}_{\mu
}U(a)^{\dagger }={Q}_{\mu }+a_{\mu }$ and $U(a)\psi (q,p)=e^{ia_{\mu
}p^{\mu }}\psi (q^{\mu }+a^{\mu },p)$, with  $[\hat{P}_{\mu },\hat{P}_{\nu }]=0$.

An equation of motion describing spin-zero particles is obtained
by using the Casimir invariant $P^{2}=m^{2}$,  leading to
 $P^{\mu }P_{\mu }\phi
(p,q)=m^{2}\phi (p,q)$, that is given explicitly by
\begin{equation}
\frac{-{\hbar}^2}{4}\frac{\partial ^{2}\phi (p,q)}{\partial q^{\mu }\partial q_{\mu }%
}-i\hbar p^{\mu }\frac{\partial \phi (p,q)}{\partial q^{\mu }}+(p^{\mu }p_{\mu
}-m^{2})\phi (p,q)=0.  \label{KGjan1}
\end{equation}%
This is a Klein-Gordon-like equation written in phase space. The field~$\phi (q,p)$ is associated with a Wigner function by~\cite{Oliveira,Amorim0}
\begin{equation}
f_{W}(q,p)=\phi (q,p)\star \phi ^{\dagger }(q,p).  \label{junho271}
\end{equation}%
This result is important to provide a full representation.

Some details for the Hilbert space in the phase space is presented in Appendix A and a  representation for spin 1/2 particles is detailed in the Appendix B. Here it is important to mention that in this representation, the free field has already the content of a non-commutative theory due to the presence of the Moyal product defining the Lie-algebra representation. This aspect is  central for providing the correct interpretation of the theory, when the connection of $\phi (q,p)$, the quasi-amplitude of probability, with the Wigner function is established, as given in eq.~(\ref{junho271}). An analysis of the gauge invariance following similar steps of the standard gauge theory are expected.

\section{Isospin  in phase space}

A local isotopic gauge transformation in phase space is such that
\begin{eqnarray}
\psi(q,p) \rightarrow \psi^{\prime}(q,p)=S^{-1} \star \psi(q,p,),
\label{eq:Y-M.1}
\end{eqnarray}%
where $S$ represents a unitary matrix $2\times2$ with $SS^{-1}=S^{-1}S=1$,
and $\psi(q,p)$ is a two-component wavefunction. In this case, $\psi(q,p)$
describes a field with isotopic spin $1/2$. The matrix $S$ is given as
\begin{eqnarray}
S=\exp(-i\boldsymbol{\tau}\cdot \boldsymbol{\alpha}),  \label{eq:Y-M.2t}
\end{eqnarray}%
where $\boldsymbol{\alpha}=(\alpha_1,\alpha_2,\alpha_3)$, and $\boldsymbol{%
\tau}=(\tau_1,\tau_2,\tau_3)$ represent the Pauli matrices.

From the gauge transformation, eq.~(\ref{eq:Y-M.1}), one has
\begin{eqnarray}
\psi(p,q)&=&S \star \psi^{\prime}(p,q).  \label{eq:Y-M.2}
\end{eqnarray}%
Therefore, by multiplying the left of eq.~(\ref{eq:Y-M.1}) by $S\star$
leads to
\begin{eqnarray}
S \star\psi^{\prime}& =&S \star S^{-1} \star \psi .  \notag
\end{eqnarray}
Expanding the star product into power series of $\hbar$, keeping terms of
first order, we obtain
\begin{eqnarray}
S \star\psi^{\prime}& =&S \star S^{-1} \star \psi  \notag \\
& =&S \left[1+ \frac{i\hbar}{2}\left( \frac{\overleftarrow{\partial}}{%
\partial q} \frac{\overrightarrow{\partial}}{\partial p} -\frac{%
\overleftarrow{\partial}}{\partial p}\frac{\overrightarrow{\partial}}{%
\partial q} \right) \right] S^{-1} \left[1+ \frac{i\hbar}{2}\left( \frac{%
\overleftarrow{\partial}}{\partial q} \frac{\overrightarrow{\partial}}{%
\partial p} -\frac{\overleftarrow{\partial}}{\partial p}\frac{%
\overrightarrow{\partial}}{\partial q} \right) \right] \psi + O(\hbar^2)
\notag \\
& =& \psi,
\end{eqnarray}
where the scalar identity is used
\begin{equation}
\frac{\partial Y^{-1}}{\partial x}= - Y^{-1} \frac{\partial Y}{\partial x}
Y^{-1}.  \label{eq:Y-M.3}
\end{equation}%
This leads to
\begin{eqnarray}
S \star S^{-1} & =&S \left[1+ \frac{i\hbar}{2}\left( \frac{\overleftarrow{%
\partial}}{\partial q} \frac{\overrightarrow{\partial}}{\partial p} -\frac{%
\overleftarrow{\partial}}{\partial p}\frac{\overrightarrow{\partial}}{%
\partial q} \right) \right] S^{-1} + O(\hbar^2)  \notag \\
& =&1.  \notag
\end{eqnarray}

In analogy with the abelian case where the covariant derivative is $D_{\mu}%
{\star}=\left( p_{\mu} \star + i e A_{\mu} \star\right) $, for isospin the definition is
\begin{equation}
G_{\mu} {\star}=  p_\mu \star + i\epsilon (\boldsymbol{\tau} \cdot
\mathbf{B}_\mu) \star   ,  \label{eq:Y-M.4}
\end{equation}
where $\mathbf{B}_\mu=(B_\mu^1,B_\mu^2,B_\mu^3)$. 
The invariance of local gauge transformation requires
\begin{equation}
G_{\mu} {\star} \psi(q,p)= S\star G_{\mu}^{^{\prime}} {\star}
\psi^{\prime}(p,q).  \label{eq:Y-M.05}
\end{equation}
Using eq.~(\ref{eq:Y-M.1}) in eq.~(\ref{eq:Y-M.05}), the following
transformation for $B_\mu$ is 
\begin{eqnarray}
G_{\mu}{\star} \psi(q,p)&=& S\star ( p_\mu \star ) S^{-1} \star \psi (p,q) + i\epsilon S\star (%
\boldsymbol{\tau} \cdot \mathbf{B}^{\prime}_\mu) \star S^{-1} \star \psi
(p,q).  \label{eq:Y-M.6}
\end{eqnarray}
The first term on the right side of eq.~(\ref{eq:Y-M.6}) is considered
separately leading to
\begin{eqnarray}
T_{9} &=& S\star ( p_\mu \star ) S^{-1} \star \psi (p,q)  \notag \\
&=& S\star \left( p_\mu S^{-1} \right) \star \psi (p,q) + \frac{i\hbar}{2}
S\star S^{-1}\star \frac{\partial S}{\partial q^{\mu}} \star S^{-1}\star
\psi (p,q) - \frac{i\hbar}{2} \frac{\partial \psi(p,q)}{\partial q^{\mu}}.
\label{eq:Y-M.6.1}
\end{eqnarray}
Here the following relation is used
\begin{equation}
\frac{\partial Y^{-1}}{\partial x}\star= -Y^{-1} \star \frac{\partial Y}{
\partial x} \star Y^{-1}\star.  \label{eq:Y-M.6.1aa}
\end{equation}
Using eq.~(\ref{eq:d.10}), the first term of the eq.~(\ref{eq:Y-M.6.1})
becomes $S\star( p_\mu S^{-1} )= p_\mu ( S\star S^{-1} ) + \frac{i\hbar}{2} (%
\frac{\partial S}{\partial q^{\mu}}) \star S^{-1} $, then
\begin{eqnarray*}  \label{eq:Y-M.6.3}
T_{9} &=& p_\mu \psi (p,q) + i\hbar \frac{\partial S}{\partial q^{\mu}} \star
S^{-1}\star \psi (p,q) - \frac{i\hbar}{2} \frac{\partial \psi(p,q)}{\partial
q^{\mu}}  \notag \\
&=& p_\mu \star \psi (p,q) + i\hbar \frac{\partial S}{\partial q^{\mu}}
\star S^{-1}\star \psi (p,q).  \notag \\
\end{eqnarray*}
Then eq.~(\ref{eq:Y-M.6}) becomes
\begin{eqnarray*}
i\epsilon (\boldsymbol{\tau} \cdot \mathbf{B}_\mu) \star \psi (p,q) &=& i\hbar \frac{\partial S}{\partial q^{\mu}} \star S^{-1}\star \psi (p,q)\
+ \epsilon S\star (\boldsymbol{\tau} \cdot \mathbf{B}^{\prime}_\mu)\star
S^{-1} \star \psi (p,q)  \notag
\end{eqnarray*}
 leading to
\begin{eqnarray*}
(\boldsymbol{\tau} \cdot \mathbf{B}_\mu) \star &=& \frac{\hbar}{\epsilon}
\frac{\partial S}{\partial q^{\mu}} \star S^{-1}\star + S\star (\boldsymbol{%
\tau} \cdot \mathbf{B}^{\prime}_\mu) \star S^{-1} \star.
\end{eqnarray*}

In order to isolate the term $(\boldsymbol{\tau} \cdot \mathbf{B}^{\prime}_\mu)$,
multiply the above equation by $S$, which gives
\begin{equation}
S\star (\boldsymbol{\tau} \cdot \mathbf{B}^{\prime}_\mu) = (\boldsymbol{\tau}
\cdot \mathbf{B}_\mu) \star S - \frac{\hbar}{\epsilon} \frac{\partial S}{%
\partial q^{\mu}},  \label{eq47}
\end{equation}
where $(\boldsymbol{\tau} \cdot \mathbf{B}^{\prime}_\mu) \star S^{-1} \star
S=(\boldsymbol{\tau} \cdot \mathbf{B}^{\prime}_\mu) $ e $\frac{\partial S}{%
\partial q^{\mu}} \star S^{-1} \star S = \frac{\partial S}{\partial q^{\mu}}$%
. Multiplying eq.~(\ref{eq47}) by $S^{-1} \star$ from the left, leads to
\begin{equation}
\boldsymbol{\tau} \cdot \mathbf{B}^{\prime}_\mu = S^{-1} \star (\boldsymbol{%
\tau} \cdot \mathbf{B}_\mu) \star S - \frac{\hbar}{\epsilon} S^{-1} \star
\frac{\partial S}{\partial q^{\mu}}.  \label{eq:Y-M.7}
\end{equation}
Therefore, the field $B_\mu$ obeys the gauge transformation given by eq.~(\ref%
{eq:Y-M.1}). In analogy with the electromagnetic field, where the relation $%
[ D_{\mu} {\star}, D_{\nu} {\star}] $ is used to find $\mathcal{F}_{\mu \nu}$%
, for the case of Isospin, this leads to
\begin{eqnarray}
[ G_{\mu} {\star}, G_{\nu} {\star} ] \psi(q,p) &=& \left[ p_{\mu} \star , p_{\nu} \star \right] \psi(q,p) + \left[ p_{\mu}
\star , i \epsilon (\boldsymbol{\tau} \cdot \mathbf{B}_\nu) \star \right]
\psi(q,p) + \left[ i \epsilon (\boldsymbol{\tau} \cdot \mathbf{B}_\mu) \star
, p_{\nu} \star \right]\psi(q,p)  \notag \\
&&+ \left[ i \epsilon (\boldsymbol{\tau} \cdot \mathbf{B}_\mu) \star , i
\epsilon (\boldsymbol{\tau} \cdot \mathbf{B}_\nu) \star \right]\psi(q,p),
\label{eq:Y-M.7.1}
\end{eqnarray}
where eq. (\ref{eq:Y-M.4}) has been used. Now each term
is calculated separately. 

The first term becomes $\left[ p_{\mu} \star , p_{\nu} \star \right] \psi(q,p)=0$.
The second term is
\begin{eqnarray}
T_{10a} &=& \left[ p_{\mu} \star , i \epsilon (\boldsymbol{\tau} \cdot
\mathbf{B}_\nu) \star \right] \psi(q,p)  \notag \\
&=& - i^2\hbar \epsilon \boldsymbol{\tau} \cdot \frac{\partial \mathbf{B}_\nu%
}{\partial q^\mu} \star \psi(q,p).  \label{eq:Y-M.7.3}
\end{eqnarray}
Using eq. (\ref{eq:d.10}) in Appendix B,   the following relation is derived,
\begin{equation}
(\boldsymbol{\tau} \cdot \mathbf{B}_\nu) \star ( p_{\mu}\psi(q,p))= p_\mu
\left( (\boldsymbol{\tau} \cdot \mathbf{B}_\nu) \star \psi (p,q) \right) +
\frac{i\hbar}{2} \boldsymbol{\tau} \cdot \frac{\partial \mathbf{B}_\nu}{%
\partial q^\mu} \star \psi (p,q).  \notag
\end{equation}

In analogy with eq.~(\ref{eq:Y-M.7.3}), the third term of eq.~(\ref%
{eq:Y-M.7.1}) takes the form
\begin{eqnarray}
T_{10b} &=& \left[ i \epsilon (\boldsymbol{\tau} \cdot \mathbf{B}_\mu )
\star , p_{\nu} \star \right]\psi(q,p)  \notag \\
&=& i^2\hbar \epsilon \boldsymbol{\tau} \cdot \frac{\partial \mathbf{B}_\mu}{%
\partial q^\nu} \star \psi(q,p) .  \label{eq:Y-M.7.5}
\end{eqnarray}

The last term is given by
\begin{eqnarray}
T_{10c} &=& \left[ i \epsilon (\boldsymbol{\tau} \cdot \mathbf{B}_\mu )
B_{\mu} \star , i \epsilon (\boldsymbol{\tau} \cdot \mathbf{B}_\nu )B_{\nu}
\star \right]\psi(q,p)  \notag \\
&=& i^2 \epsilon (\boldsymbol{\tau} \cdot \mathbf{B}_\mu ) \star \epsilon (%
\boldsymbol{\tau} \cdot \mathbf{B}_\nu ) \star \psi(q,p) - i^2 \epsilon (%
\boldsymbol{\tau} \cdot \mathbf{B}_\nu )\star \epsilon (\boldsymbol{\tau}
\cdot \mathbf{B}_\mu ) \star \psi(q,p) .  \label{eq:Y-M.7.6}
\end{eqnarray}

Hence,  eq. (\ref{eq:Y-M.7.1}) becomes
\begin{eqnarray}
[ G_{\mu} {\star}, G_{\nu} {\star}] \psi(q,p) &=& \hbar \epsilon \boldsymbol{%
\tau} \cdot \left( \frac{\partial \mathbf{B}_{\nu} }{\partial q^{\mu}} -
\frac{\partial \mathbf{B}_{\mu} }{\partial q^{\nu}} \right) \star \psi(q,p)
+   \epsilon^2 \left\{ (\boldsymbol{\tau} \cdot \mathbf{B}_\mu ), (%
\boldsymbol{\tau} \cdot \mathbf{B}_\nu )\right\}_M \star \psi(q,p).
\end{eqnarray}
As a consequence, a tensor strength  for the isospin is defined as
\begin{equation}
\mathcal{K}_{\mu \nu} = \frac{\partial b_{\mu} }{\partial q^{\nu}} - \frac{%
\partial b_{\nu} }{\partial q^{\mu}} +\epsilon \left\{
b_{\mu},b_{\nu}\right\}_M,  \label{eq:Y-M.8}
\end{equation}
where $b_\mu=\boldsymbol{\tau} \cdot \mathbf{B}_\mu$.
For the Isospin, using eq.~(\ref{eq:Y-M.7}) the tensor is transformed to the
form
\begin{equation}
\mathcal{K}_{\mu \nu}^{^{\prime}} = S^{-1} \star \mathcal{K}_{\mu \nu} \star
S .  \label{eq:Y-M.9}
\end{equation}
Considering
\begin{eqnarray*}
\mathcal{K}_{\mu \nu} \star \psi &=& S \star \mathcal{K}_{\mu
\nu}^{^{\prime}} \star \psi^{\prime}  \notag \\
&=& S \star \mathcal{K}_{\mu \nu}^{^{\prime}} \star S^{-1} \star \psi,  \notag
\end{eqnarray*}
then $\mathcal{K}_{\mu \nu} \star$ has the form
\begin{eqnarray}
\mathcal{K}_{\mu \nu} \star &=& S \star \mathcal{K}_{\mu \nu}^{^{\prime}}
\star S^{-1} \star .  \notag
\end{eqnarray}
Multiplying by $S$ on the right-hand side leads to
\begin{eqnarray}
\mathcal{K}_{\mu \nu} \star S=S\star \mathcal{K}_{\mu \nu}^{^{\prime}}.
\notag
\end{eqnarray}
By multiplying by $S^{-1} \star$ on the left-hand side leads to
\begin{eqnarray}
\mathcal{K}_{\mu \nu}^{^{\prime}} &=& S^{-1} \star \mathcal{K}_{\mu
\nu}\star S.  \notag
\end{eqnarray}
Then eq.~(\ref{eq:Y-M.8}) becomes
\begin{eqnarray}
\mathcal{K}_{\mu \nu}& =&\boldsymbol{\tau} \cdot \frac{\partial \mathbf{B}_\mu }{\partial q^{\nu}}
- \boldsymbol{\tau} \cdot \frac{\partial \mathbf{B}_\nu }{\partial q^{\mu}}
 +\epsilon \left( (\boldsymbol{\tau} \cdot \mathbf{B}%
_\mu)\star (\boldsymbol{\tau} \cdot \mathbf{B}_\nu) - (\boldsymbol{\tau}
\cdot \mathbf{B}_\nu) \star (\boldsymbol{\tau} \cdot \mathbf{B}_\mu)\right).
\label{eq:Y-M.14}
\end{eqnarray}

The Pauli matrices have the property
\begin{equation}
(\mathbf{A}\cdot \boldsymbol{\sigma})(\mathbf{B}\cdot \boldsymbol{\sigma})=
\mathbf{A} \cdot \mathbf{B}+ i \boldsymbol{\sigma} \cdot (\mathbf{A} \times
\mathbf{B}).  \label{eq:by.8a}
\end{equation}%
A similar property is connected to the star product
\begin{equation}
(\mathbf{A}\cdot \boldsymbol{\sigma}) \star (\mathbf{B}\cdot \boldsymbol{%
\sigma})=\tilde{A} \star \tilde{B},  \label{eq:by.8b}
\end{equation}%
where $\mathbf{A}\cdot \boldsymbol{\sigma}$ is given by
\begin{equation*}
\mathbf{A}\cdot \boldsymbol{\sigma}= A_x \sigma_y+ A_y \sigma_x+ A_z
\sigma_z = \left(%
\begin{array}{ccc}
A_z & A_x - i A_y &  \\
A_x + i A_y & - A_z &
\end{array}
\right)=\tilde{A},
\end{equation*}
and the term $(\mathbf{B}\cdot \boldsymbol{\sigma})$, has the form
\begin{equation*}
\mathbf{B}\cdot \boldsymbol{\sigma}= B_x \sigma_y+ B_y \sigma_x+ B_z
\sigma_z = \left(%
\begin{array}{ccc}
B_z & B_x - i B_y &  \\
B_x + i B_y & - B_z &
\end{array}
\right)=\tilde{B}.
\end{equation*}
Expanding the star product of eq.~(\ref{eq:by.8b}) leads to
\begin{eqnarray}
\tilde{A} \star \tilde{B}&=& \tilde{A} \tilde{B} + \frac{i\hbar}{2} \frac{\partial \tilde{A}}{
\partial q}\frac{\partial \tilde{B} }{ \partial p} - \frac{i\hbar}{2} \frac{%
\partial \tilde{A}}{ \partial p}\frac{\partial \tilde{B} }{ \partial q} +
O(\hbar^2).  \label{eq:by.8c}
\end{eqnarray}%
The first term is written as
\begin{equation}  \label{eqn:linear_ssK3}
\begin{matrix}
&  & \tilde{A} \tilde{B} = \left(%
\begin{array}{ccc}
A_z & A_x - i A_y &  \\
A_x + i A_y & - A_z &
\end{array}
\right) \left(%
\begin{array}{ccc}
B_z & B_x - i B_y &  \\
B_x + i B_y & - B_z &
\end{array}
\right) \\
&  & = \mathbf{A}\cdot \mathbf{B} \mathbf{1} + i ( A_x B_y - A_y B_x)_{\hat{z%
}} \sigma_z + i(A_zB_x -A_x B_z)_{\hat{y}} \sigma_y + i(A_yB_z -A_z B_y)_{%
\hat{x}} \sigma_x.
\end{matrix}%
  \notag
\end{equation}
And the second and third terms in eq.~(\ref{eq:by.8c}) take the form
\begin{eqnarray}
\frac{i\hbar}{2} \frac{\partial \tilde{A}}{ \partial q}\frac{\partial \tilde{B}
}{ \partial p} &=& \frac{i\hbar}{2} \frac{\partial \mathbf{A} }{ \partial q}%
\frac{\partial \mathbf{B} }{ \partial p} \mathbf{1} + \frac{i\hbar}{2} i
\left( \frac{\partial A_x }{ \partial q_x}\frac{\partial B_y }{ \partial p_y}%
- \frac{\partial A_y }{ \partial q_y} \frac{\partial B_x }{ \partial p_x}
\right)_{\hat{z}} \sigma_z \\
&+& \frac{i\hbar}{2} \left( \frac{\partial A_z }{ \partial q_z}\frac{%
\partial B_x }{ \partial p_x} -\frac{\partial A_x }{ \partial q_x}\frac{%
\partial B_z }{ \partial p_z} \right)_{\hat{y}} \sigma_y +\frac{i\hbar}{2} i
\left( \frac{\partial A_y }{ \partial q_y}\frac{\partial B_z }{ \partial p_z}%
- \frac{\partial A_z }{ \partial q_z} \frac{\partial B_y }{ \partial p_y}
\right)_{\hat{x}} \sigma_x \\
-\frac{i\hbar}{2} \frac{\partial \tilde{A}}{ \partial p}\frac{\partial \tilde{B}
}{ \partial q} &=&- \frac{i\hbar}{2} \frac{\partial \mathbf{A} }{ \partial p}%
\frac{\partial \mathbf{B} }{ \partial q} \mathbf{1} - \frac{i\hbar}{2} i
\left( \frac{\partial A_x }{ \partial p_x}\frac{\partial B_y }{ \partial q_y}%
- \frac{\partial A_y }{ \partial p_y} \frac{\partial B_x }{ \partial q_x}
\right)_{\hat{z}} \sigma_z  \notag \\
&+& \frac{i\hbar}{2} \left( \frac{\partial A_z }{ \partial p_z}\frac{%
\partial B_x }{ \partial q_x} -\frac{\partial A_x }{ \partial p_x}\frac{%
\partial B_z }{ \partial q_z} \right)_{\hat{y}} \sigma_y -\frac{i\hbar}{2} i
\left( \frac{\partial A_y }{ \partial p_y}\frac{\partial B_z }{ \partial q_z}%
- \frac{\partial A_z }{ \partial p_z} \frac{\partial B_y }{ \partial q_y}
\right)_{\hat{x}} \sigma_x .
\;\;\;\;\;\;\;\;\;\;\;\;\;\;\;\;\;\;\;\;\;\;\;\;\;\;\;\;\;\;\;\;\;\;\;\;\;\;%
\;\;\;\;\;\;\;\;\;\;\;\;\;\;\;\;\;\;\;\;\;\;\;\;\;\;\;\;\;\;  \notag
\end{eqnarray}
Therefore, using these results, eq.~(\ref{eq:by.8c}) becomes
\begin{eqnarray}
\tilde{A} \star \tilde{B}&=& \mathbf{A} \star \mathbf{B} + i \boldsymbol{\sigma} \cdot \left( \mathbf{%
A} \times \mathbf{B} \right)_{\star} ,  \notag
\end{eqnarray}%
where $\left( \mathbf{A} \times \mathbf{B} \right)_{\star \hat{k}} =
\left(A_i \star B_j - A_j \star B_i \right)_{\hat{k}}, \notag
$ with $\hat{i}, \hat{j}, \hat{k}=\hat{x} , \hat{y}, \hat{z}.$

The terms of eq.~(\ref{eq:Y-M.14}) become
\begin{eqnarray}
( \boldsymbol{\tau}\cdot \mathbf{B}_{\mu}) \star (\boldsymbol{\tau}\cdot
\mathbf{B}_{\nu}) &=& \mathbf{B}_{\mu} \star \mathbf{B}_{\nu} + i
\boldsymbol{\tau}\cdot \left( \mathbf{B}_{\mu} \times \mathbf{B}_{\nu}
\right)_{\star},  \notag
\end{eqnarray}%
and%
\begin{eqnarray}
-( \boldsymbol{\tau}\cdot \mathbf{B}_{\nu}) \star (\boldsymbol{\tau}\cdot
\mathbf{B}_{\mu}) &=& - \mathbf{B}_{\nu} \star \mathbf{B}_{\mu} - i
\boldsymbol{\tau}\cdot \left( \mathbf{B}_{\nu} \times \mathbf{B}_{\mu}
\right)_{\star}.  \notag
\end{eqnarray}%
Then eq.~(\ref{eq:Y-M.14}) is written as
\begin{eqnarray}
\mathcal{K}_{\mu \nu} & =&\boldsymbol{\tau} \cdot \mathbf{k}_{\mu \nu} + \epsilon \left\{ \mathbf{B%
}_\mu , \mathbf{B}_\nu \right\}_{M},  \label{eq:Y-M.15}
\end{eqnarray}
where
\begin{eqnarray}
\mathbf{k}_{\mu \nu} & =& \frac{\partial \mathbf{B}_\mu }{\partial q^{\nu}}
- \frac{\partial \mathbf{B}_\nu }{\partial q^{\mu}} + i \epsilon \left(
\mathbf{B}_{\mu} \times \mathbf{B}_{\nu} \right)_{\star} - i \epsilon \left(
\mathbf{B}_{\nu} \times \mathbf{B}_{\mu} \right)_{\star}.
\label{eq:Y-M.15sa}
\end{eqnarray}
The infinitesimal gauge transformation for eq.~(\ref{eq:Y-M.2t}) is given by
\begin{eqnarray}
S=exp(-i\boldsymbol{\tau}\cdot \boldsymbol{\alpha})= 1-i\boldsymbol{\tau}%
\cdot \boldsymbol{\alpha}.  \label{eq:Y-M.17}
\end{eqnarray}%
Then eq.~(\ref{eq:Y-M.7}) becomes
\begin{eqnarray}
\boldsymbol{\tau}\cdot \mathbf{B^{\prime}_\mu} &=&(1 + i\boldsymbol{\tau}\cdot \boldsymbol{\alpha} )\star \left(
\boldsymbol{\tau}\cdot\mathbf{B_\mu}\right) \star (1 -i\boldsymbol{\tau}%
\cdot \boldsymbol{\alpha} ) -\frac{1}{\epsilon}\left[( 1+i\boldsymbol{\tau}%
\cdot \boldsymbol{\alpha})\star \frac{\partial (1-i\boldsymbol{\tau}\cdot
\boldsymbol{\alpha})}{ \partial q^\mu}\right]  \notag \\
&=& \boldsymbol{\tau}\cdot\mathbf{B_\mu} + \boldsymbol{\tau}\cdot(\mathbf{%
B_\mu} \times \boldsymbol{\alpha})_{\star} - \boldsymbol{\tau} \cdot(%
\boldsymbol{\alpha} \times \mathbf{B_\mu})_{\star} +\frac{i}{\epsilon}%
\boldsymbol{\tau}\cdot\frac{\partial ( \boldsymbol{\alpha} ) }{ \partial
q^\mu} - i \left\{ \mathbf{B_\mu}, \boldsymbol{\alpha}\right\}_M.
\label{eq:Y-M.17aa}
\end{eqnarray}

Assuming a field with isotopic spin $1/2$,  the final Lagrangian is
\begin{eqnarray}
\mathcal{L}_{total}&=&-\frac{1}{2} \left[ \bar{\psi}(p,q)\gamma^{\mu } \star \left( p_{\mu}
\star \psi (p,q) + i\epsilon ( \boldsymbol{\tau}\cdot\mathbf{B_\mu} ) \star
\psi (p,q) \right) \right]  \notag \\
&& + \left[\left(\bar{\psi}(p,q)\star p_{\mu}+ i\epsilon \bar{\psi}(p,q)
\star ( \boldsymbol{\tau}\cdot\mathbf{B_\mu}) \right)\star \gamma^{\mu }
\psi (p,q) \right]  \notag \\
&&- m \bar{\psi}(p,q) \star \psi (p,q) + 2 p_{\mu } \bar{\psi}(p,q)
\gamma^{\mu } \star \psi (p,q) -\frac{1}{4} \mathcal{K}^{\mu\nu}\mathcal{K}%
_{\mu \nu},    \label{eq:d.25}
\end{eqnarray}
 describing an isospin in phase space.

As in the previous sections, the star product is expanded in power series of $\hbar$ up to
first order. Then expanding eq.~(\ref{eq:Y-M.1}), the isotopic gauge transformation is
given by
\begin{eqnarray}
\psi(q,p) \rightarrow \psi^{\prime}(q,p)=S^{-1} \psi(q,p).  \label{eq:YY-M.1}
\end{eqnarray}%
The covariant derivative is defined as
\begin{equation}
G_\mu= \left( p_\mu- \frac{i\hbar}{2} \frac{\partial }{ \partial q^\mu} + \frac{i
}{2}\epsilon \boldsymbol{\tau}\cdot\mathbf{B_\mu} \right) .
\label{eq:YY-M.2}
\end{equation}
Requiring that the covariant derivative is over the local gauge leads to
\begin{eqnarray}
G_\mu \psi(p,q)&=&S G^{\prime}_\mu \psi^{\prime}(p,q)  \notag \\
&=&S\left( p_\mu- \frac{i\hbar}{2} \frac{\partial }{ \partial q^\mu} + \frac{i}{2}%
\epsilon \boldsymbol{\tau}\cdot\mathbf{B^{\prime}_\mu} \right)
\psi^{\prime}(p,q)  \notag \\
&=& p_\mu \psi (p,q) + \frac{i\hbar}{2} \frac{\partial S }{ \partial q^\mu}
S^{-1} \psi (p,q) - \frac{i\hbar}{2} \frac{\partial \psi (p,q) }{ \partial q^\mu}
+ \frac{i}{2} \epsilon S \boldsymbol{\tau}\cdot\mathbf{B^{\prime}_\mu}
S^{-1}\psi (p,q).  \notag
\end{eqnarray}
Therefore, the isotope field, $\mathbf{B^{\prime}_\mu}$, transforms as
\begin{equation}
\boldsymbol{\tau}\cdot\mathbf{B^{\prime}_\mu} = S^{-1} \boldsymbol{\tau}\cdot%
\mathbf{B_\mu} S - \frac{1}{\epsilon}S^{-1} \frac{\partial S }{ \partial
q^\mu},  \label{eq:YY-M.3}
\end{equation}
where the identity in eq.~(\ref{eq:Y-M.3}) is used.

Similar to the
non-abelian gauge field $F_{\mu \nu }$, for the isospin field, eq.~(\ref%
{eq:Y-M.15}), becomes
\begin{eqnarray}
\mathcal{K}_{\mu \nu} & =&\boldsymbol{\tau} \cdot \frac{\partial \mathbf{B}%
_\mu }{\partial q^{\nu}} - \boldsymbol{\tau} \cdot \frac{\partial \mathbf{B}%
_\nu }{\partial q^{\mu}} + i \epsilon \boldsymbol{\tau}\cdot \left( \mathbf{B%
}_{\mu} \times \mathbf{B}_{\nu} \right)_{\star} - i \epsilon \boldsymbol{\tau%
}\cdot \left( \mathbf{B}_{\nu} \times \mathbf{B}_{\mu} \right)_{\star}
\notag \\
&& + \epsilon \left\{ \mathbf{B}_\mu , \mathbf{B}_\nu \right\}_{M},
\label{eq:Y-M.15k}
\end{eqnarray}
where
\begin{eqnarray*}
\left( \mathbf{B}_{\mu} \times \mathbf{B}_{\nu}\right)_{\star}&=& \left(
B_{\mu}^{x} \star B_{\nu}^{y}- B_{\mu}^{y} \star B_{\nu}^{x}\right)_{\hat{z}%
} +\left( B_{\mu}^{z} \star B_{\nu}^{x}- B_{\mu}^{x} \star
B_{\nu}^{z}\right)_{\hat{y}}  \notag \\
&&+\left( B_{\mu}^{y} \star B_{\nu}^{z}- B_{\mu}^{z} \star
B_{\nu}^{y}\right)_{\hat{x}}.
\end{eqnarray*}
Expanding the star product in power series of $\hbar$ in the zeroth order $%
\star=\exp \left[ \frac{i\hbar }{2}\left( \frac{\overleftarrow{\partial }}{%
\partial q}\frac{\overrightarrow{\partial }}{\partial p}-\frac{%
\overleftarrow{\partial }}{\partial p}\frac{\overrightarrow{\partial }}{%
\partial q}\right) \right]=1$. Then this equation is written as
\begin{eqnarray*}
\left( \mathbf{B}_{\mu} \times \mathbf{B}_{\nu}\right)&=& \left( B_{\mu}^{x}
B_{\nu}^{y}- B_{\mu}^{y} B_{\nu}^{x}\right)_{\hat{z}} +\left( B_{\mu}^{z}
B_{\nu}^{x}- B_{\mu}^{x} B_{\nu}^{z}\right)_{\hat{y}}  \notag \\
&&+\left( B_{\mu}^{y} B_{\nu}^{z}- B_{\mu}^{z} B_{\nu}^{y}\right)_{\hat{x}}.
\end{eqnarray*}
Therefore, eq.~(\ref{eq:Y-M.15k}) becomes
\begin{eqnarray}
\mathcal{K}_{\mu \nu} &=&\boldsymbol{\tau} \cdot \bm{k}_{\mu \nu},  \label{eq:Y-M.15k1}
\end{eqnarray}
where
\begin{eqnarray}
\bm{k}_{\mu \nu}= \frac{\partial \mathbf{B}_\mu }{\partial q^{\nu}} - \frac{%
\partial \mathbf{B}_\nu }{\partial q^{\mu}} + i \epsilon 2 \left( \mathbf{B}%
_{\mu} \times \mathbf{B}_{\nu} \right).  \label{eq:Y-M.15k11}
\end{eqnarray}
Using eq. (\ref{eq:YY-M.1}), this becomes
\begin{equation*}
\bm{k}^{\prime}_{\mu \nu} = S^{-1}\bm{k}_{\mu \nu}S.  \label{eq:Y-M.5}
\end{equation*}
The field $\mathbf{B_\mu}$ in eq.~(\ref{eq:YY-M.3}), under a gauge
transformation in infinitesimal form, eq. (\ref{eq:Y-M.17}), is
\begin{eqnarray}
\boldsymbol{\tau}\cdot\mathbf{B^{\prime}_\mu}&=&(1+i\boldsymbol{\tau}\cdot \boldsymbol{\alpha}) \boldsymbol{\tau}\cdot%
\mathbf{B_\mu}(1-i\boldsymbol{\tau}\cdot \boldsymbol{\alpha})- \frac{1}{%
\epsilon}\left[(1+i\boldsymbol{\tau}\cdot \boldsymbol{\alpha}) \frac{%
\partial (1-i\boldsymbol{\tau}\cdot \boldsymbol{\alpha}) }{ \partial q^\mu}%
\right]  \notag \\
&=&\boldsymbol{\tau}\cdot \mathbf{B_\mu} +2\boldsymbol{\tau}\cdot (\mathbf{%
B_\mu}\times \boldsymbol{\alpha}) +\frac{i}{\epsilon}\boldsymbol{\tau}\cdot%
\frac{\partial (\boldsymbol{\alpha}) }{ \partial q^\mu}.  \notag
\end{eqnarray}
Then the infinitesimal transformation of $\mathbf{B_\mu}$ is
\begin{eqnarray}
\mathbf{B^{\prime}_\mu}= \mathbf{B_\mu} +2 (\mathbf{B_\mu}\times \boldsymbol{%
\alpha}) +\frac{i}{\epsilon} \frac{\partial (\boldsymbol{\alpha}) }{
\partial q^\mu}.  \notag
\end{eqnarray}
Using $\bm{k}_{\mu\nu}$, the Lagrangian density is written as
\begin{equation}
\mathcal{L}_k=-\frac{1}{4}\bm{k}^{\mu\nu}\cdot \bm{k}_{\mu\nu}.  \notag
\end{equation}

The final Lagrangian density for isotopic spin $1/2$ is
\begin{eqnarray}
\mathcal{L}_{final} &=&-\frac{1}{2} \bigg[ \bar{\psi}(p,q)\gamma^{\mu }
\left( G_\mu \psi (p,q)\right) + \left( \bar{G_\mu} \bar{\psi} (p,q) \right)
\gamma^{\mu } \psi (p,q) \bigg]  \notag \\
&-& m \bar{\psi}(p,q) \psi (p,q) + 2 p_{\mu } \bar{\psi}(p,q) \gamma^{\mu }
\psi (p,q) -\frac{1}{4}\bm{k}^{\mu\nu}\cdot \bm{k}_{\mu\nu},
\label{eq:Y-M.fin}
\end{eqnarray}
where $G_\mu$ is given by eq.~(\ref{eq:YY-M.2}) and $\bar{G_\mu}$ is
\begin{equation*}
\bar{G_\mu}= p_\mu- \frac{i\hbar}{2} \frac{\partial }{ \partial q^\mu} + \frac{i}{%
2}\epsilon \boldsymbol{\tau}\cdot\mathbf{B_\mu} .
\end{equation*}
This result is equivalent to the usual Lagrangian as obtained for Yang-Mills gauge
theory. However, for $\epsilon=0$, the covariant derivative reduces, consistently, to the momentum operator in the phase space representation, i.e. $P=p\star$. In the following section these results are used to consider a nucleon in an external field. These results are a test for the consistency of the representation.

\section{Nucleon in an External Field}

Consider a nucleon in an external field that is intense enough to discard the gauge field self interaction in zero order (natural units are used). Then the Lagrangian in eq.~(\ref{eq:Y-M.fin}) leads to the Dirac equation
\begin{equation}
(\gamma^\mu\hat{G}_\mu-m)\psi=0,
\end{equation}
where $\hat{G}_\mu$ is given by eq. (\ref{eq:YY-M.2}).

Taking   $\mathbf{B_\mu}=(0,0,B_\mu)$, then
$$\left[\gamma^\mu\left( p_\mu- \frac{i}{2} \frac{\partial }{ \partial q^\mu} + \frac{1
}{2}\,\alpha\epsilon B_\mu\star\right)-m\right]\psi=0\,,$$ with $\alpha=\pm 1$. Except for $\alpha$ this equation is similar to that obtained for the Landau problem in phase space~\cite{diracphasespace}. Writing
\begin{equation}
B^{i}\star=\frac{1}{2}\epsilon ^{ijk}H_{j}q_{k}\star,~\ i=(1,2,3),
\end{equation}%
 yields
$\mathbf{H}%
=\left( 0,0,H\right) $.
Then the field equation is given by
\begin{equation}
\left[ \gamma ^{\mu }\gamma ^{\nu }\left( p_{\mu }\star-\frac{1}{2}\alpha\epsilon B_{\mu
}\star\right) \left( p_{\nu }\star-\frac{1}{2}\alpha\epsilon B_{\nu }\star\right) -m^{2}\right] \psi =0,
\label{ediracc}
\end{equation}%
which explicitly reads
\begin{align}
& \left( \varepsilon^{2}+i\varepsilon\frac{\partial }{\partial \tau}-\frac{1}{4}\frac{\partial ^{2}%
}{\partial \tau^{2}}\right) \psi -\Biggl\{p_{x}^{2}+p_{y}^{2}-\frac{1}{4}%
\left( \frac{\partial ^{2}}{\partial x^{2}}+\frac{\partial ^{2}}{\partial
y^{2}}\right) \notag \\ &-\frac{1}{2}\alpha\epsilon H\left[ \frac{i}{2}\left( p_{y}\frac{\partial }{\partial
p_{x}}-p_{x}\frac{\partial }{\partial p_{y}}\right) +\frac{1}{4}\left( \frac{%
\partial ^{2}}{\partial y\partial p_{x}}-\frac{\partial ^{2}}{\partial
x\partial p_{y}}\right) \right]   \notag \\
& -i\left( p_{y}\frac{\partial }{\partial y}-p_{x}\frac{\partial }{\partial x%
}\right) -\frac{1}{2}\alpha\epsilon H\left[ \left( xp_{y}-yp_{x}\right) -\frac{i}{2}\left( x\frac{%
\partial }{\partial y}-y\frac{\partial }{\partial x}\right) \right]  \notag \\
&+\frac{%
\left(\alpha\epsilon\right)^{2}H^{2}}{16}\left[ \left( x+\frac{i}{2}\frac{\partial }{\partial p_{x}}%
\right) ^{2}+\left( y+\frac{i}{2}\frac{\partial }{\partial p_{y}}\right) ^{2}%
\right] -i \frac{1}{2}\alpha\epsilon H\sigma ^{12}+m^{2}\Biggr\}\psi =0,  \label{dracph}
\end{align}%
where $\varepsilon=p_0$ and $\tau=x_0$.  For separated variables then two equations are obtained. The first one is given by
\begin{equation}
\left( -\varepsilon^{2}-i\varepsilon\frac{\partial }{\partial \tau}+\frac{1}{4}\frac{\partial ^{2}}{%
\partial \tau^{2}}\right) \varphi =-E ^{2}\varphi ,  \label{solE}
\end{equation}%
and the second one reads
\begin{align}
& \Biggl\{p_{x}^{2}+p_{y}^{2}-\frac{1}{4}\left( \frac{\partial ^{2}}{%
\partial x^{2}}+\frac{\partial ^{2}}{\partial y^{2}}\right) -i\left( p_{y}%
\frac{\partial }{\partial y}-p_{x}\frac{\partial }{\partial x}\right)  \notag
\\
& -\frac{1}{2}\alpha\epsilon H\Big[\left( xp_{y}-yp_{x}\right) +\frac{i}{2}\left( p_{y}\frac{\partial
}{\partial p_{x}}-p_{x}\frac{\partial }{\partial p_{y}}\right)  \notag \\
& -\frac{i}{2}\left( x\frac{\partial }{\partial y}-y\frac{\partial }{%
\partial x}\right) +\frac{1}{4}\left( \frac{\partial ^{2}}{\partial
y\partial p_{x}}-\frac{\partial ^{2}}{\partial x\partial p_{y}}\right) \Big]%
-i \frac{1}{2}\alpha\epsilon H\sigma ^{12}  \notag \\
& +\frac{(\alpha\epsilon)^{2}H^{2}}{16}\left[ \left( x+\frac{i}{2}\frac{\partial }{\partial
p_{x}}\right) ^{2}+\left( y+\frac{i}{2}\frac{\partial }{\partial p_{y}}%
\right) ^{2}\right] \Biggr\}\phi = E^{2}\phi .  \label{solxp}
\end{align}%

Introducing the following variable
\begin{equation}
z=p_{x}^{2}+p_{y}^{2}+ \frac{1}{2}\alpha\epsilon H\left( yp_{x}-xp_{y}\right) +\frac{(\alpha\epsilon)^{2}H^{2}}{16}%
\left( x^{2}+y^{2}\right) ,
\end{equation}%
and using
$i\sigma ^{12}\phi =-s\phi $, with $s=\pm 1$, $\omega =2z/(\alpha\epsilon H)$ and $\phi =\exp {(-\omega )}F(\omega )$, the equation for $%
F(\omega )$ is
\begin{equation}
\omega F^{^{\prime \prime }}+(1-2\omega )F^{^{\prime }}-(1-k)F=0,
\label{hyper}
\end{equation}%
where $k=\frac{2E^2}{\alpha\epsilon}+s$. This equation leads to
\begin{equation}
E^2-m^2= \frac{1}{2}\alpha\epsilon H(2n+1-s).  \label{engy2}
\end{equation}%
For the negative value of $\alpha$, the following condition holds
$$m^2> \frac{1}{2}\epsilon H( 2n+1-s) \,.$$
This implies that both states defined by $\alpha$ are physically acceptable. This result expresses the consistency of this representation in phase space. It is important to emphasize that, when a similar result was derived for the electron in an external field~\cite{diracphasespace}, the gauge field in phase space was introduced heuristically.  This is not the case here, where the principle of gauge symmetry has led to the derivation of the form for the gauge field.

\section{Conclusions}

A non-commutative-like non-abelian gauge theory is constructed in phase space. This corresponds to a realization of the Seiberg-Witten gauge theory for non-commutative fields that including non-abelian symmetries. The starting point is a Lagrangian density for the free Dirac field in phase space. This leads to a SU(2)- gauge group theory derived in phase space, and associated with the Wigner function. The formalism is applied to study the nucleon in a external isospin gauge field. This corresponds to a Landau problem for isospin, such that the Landau levels are consistently derived for the dublet isospin representation. A similar theory for a generalized non-abelian gauge group will be presented later.

\vspace{1cm}

\textbf{Acknowledgements:} This work was partially supported by FAP-DF  of Brazil. The work by AES and AFS is supported by CNPq of Brazil.


\appendix
\section{Symplectic field theory}

In order to construct a Hilbert space in $C^{\infty }(\Gamma )$, let $%
\mathcal{H}(\Gamma )$ be a linear subspace of the measurable
functions space $\psi :\Gamma \rightarrow \mathbb{C}$ which are square integrable,
i.e.
\begin{equation}
\int_{\Gamma }d^{4}pd^{4}q\psi ^{\ast }(q,p)\psi (q,p)<\infty .
\end{equation}%
A Hilbert space is introduced by defining the inner product, $\langle \cdot
|\cdot \rangle $, on $\mathcal{H}(\Gamma )$, as $\langle \psi _{1}|\psi
_{2}\rangle =\int_{\Gamma }\psi _{1}^{\ast }(q,p)\psi _{2}(q,p)d^{4}pd^{4}q,$
where $(q,p)=(q^{\mu },p^{\mu })$ and $\psi (q,p)$ is defined in $C^{\infty
}(\Gamma )$.

Consider the linear mappings $\overline{Q},\overline{P}: \mathcal{H}(\Gamma
)\rightarrow \mathcal{H}(\Gamma )$, such that $\overline{Q}%
\psi(q,p)=q\psi(q,p)$ and $\overline{P}\psi(q,p)=p\psi(q,p)$. Since $[%
\overline{Q},\overline{P}]=0$, the spectrum of $\overline{Q}$ and $\overline{%
P}$ is used to construct a basis in $\mathcal{H}(\Gamma )$, $%
\{|q,p\rangle\}$, such that
\begin{equation}
\overline{Q}|q,p\rangle=q|q,p\rangle,\ \ \ \ \overline{P}|q,p\rangle
=p|q,p\rangle.  \label{dav11}
\end{equation}
In this case, $\psi (q,p)=\langle q,p|\psi \rangle $, such that
\begin{equation*}
\int d^{4}pd^{4}q|q,p\rangle\langle q,p|=1\
\end{equation*}
and $\langle q,p\left\vert q^{\prime},p^{\prime}\right\rangle =\delta
(q-q^{\prime})\delta(p-p^{\prime})$. It is important to emphasize that
the operators $\overline{Q}$ and $\overline{P}$ are not the usual quantum
mechanics operators for position and momentum. Their physical meaning will
be considered latter. For physical interpretation, the state of a system
is described by functions $\psi(q,p)$, with the normalization condition
\begin{equation}
\langle\psi|\psi\rangle=\int d^{4}pd^{4}q\psi^{\ast}(q,p)\psi(q,p)=1.
\label{vict2}
\end{equation}

This symplectic Hilbert space, $\mathcal{H}(\Gamma )$, is taken as the
representation space of the Poincar\'{e} symmetry. Consider a unitary transformation in $\mathcal{H}(\Gamma )$ which is a
linear mapping $U:\mathcal{H}(\Gamma )\rightarrow \mathcal{H}(\Gamma )$,
where $\langle \psi _{1}|\psi _{2}\rangle $ is invariant.

\section{Abelian gauge theory for the sympletic Dirac field}

Here some aspects of the calculation for the abelian  gauge-field in phase space are presented. The objective of these results is twofold: first, the work is self contained; and second,  many steps (not presented in previous publications~\cite{Amorim0}) are important, since these are used in the more intricate case of the non-abelian symmetry as discussed in Section 3.

The Lagrangian density for the Dirac equation in
phase space is~\cite{Amorim0}
\begin{equation}
\mathcal{L}=-\frac{i}{4}\left( \frac{\partial \bar{\psi}(p,q)}{\partial
q^{\mu }}\gamma ^{\mu }\psi (q,p)-\bar{\psi}(p,q)\gamma ^{\mu }\frac{%
\partial \psi (p,q)}{\partial q^{\mu }}\right) -\bar{\psi}(p,q)\left(
m-\gamma ^{\mu }p_{\mu }\right) \psi (p,q).
\end{equation}%
This is re-written as
\begin{eqnarray}
\mathcal{L} &=&-\frac{1}{2}\left[ \bar{\psi}(p,q)\gamma ^{\mu }\left( p_{\mu
}\star \psi (p,q)\right) +\left( \bar{\psi}(p,q)\star p_{\mu }\right) \gamma
^{\mu }\psi (p,q)\right]  \notag \\
&&-m\bar{\psi}(p,q)\psi (p,q)+2p_{\mu }\bar{\psi}(p,q)\gamma ^{\mu }\psi
(p,q).  \label{eq:d.4}
\end{eqnarray}

Our goal is to analyse the invariance of eq.~(\ref{eq:d.4}) under global and
local gauge transformations. First, the analysis is developed in a general way, i.e. without defining the gauge for the system.
Using the property of star product, we have
\begin{equation}
\int f \star g dq dp = \int f g dq dp.  \label{eq:d.4.1}
\end{equation}
Then eq.(\ref{eq:d.4}) is
\begin{eqnarray}
S&=&\int \mathcal{L}\, d^4p \, d^4q  \notag \\
&=&\int\bigg[-\frac{1}{2} \left[ \bar{\psi}(p,q)\gamma^{\mu } \star \left(
p_\mu \star \psi (p,q) \right) + \left(\bar{\psi}(p,q)\star p_\mu
\right)\star \gamma^{\mu } \psi (p,q) \right]  \notag \\
&-& m \bar{\psi}(p,q) \star \psi (p,q) + 2 p_{\mu } \bar{\psi}(p,q)
\gamma^{\mu } \star \psi (p,q)\bigg] d^4p d^4q,  \label{eq:d.5}
\end{eqnarray}
where the star product is introduced. Defining the gauge
transformation in the general form leads to
\begin{eqnarray}
\psi(q,p) \rightarrow e^{-i \lambda } \star \psi(q,p),  \label{eq:d.6}
\end{eqnarray}%
\begin{eqnarray}
\bar{\psi} (p,q) \rightarrow \bar{\psi} (p,q) \star e^{i \lambda}.
\label{eq:d.7}
\end{eqnarray}%
In infinitesimal form, eq.~(\ref{eq:d.6}) takes the form
\begin{eqnarray}
\psi(q,p) \rightarrow \psi^{\prime}(q,p) &=& e^{-i \lambda } \star \psi(q,p)
\notag \\
&=& \psi(q,p) -i \lambda \star \psi(q,p)  \notag \\
\delta \psi(q,p) &=& -i \lambda \star \psi(q,p),  \notag
\end{eqnarray}%
where $\psi^{\prime}(q,p)- \psi(q,p)=\delta \psi(q,p) $. Then the gauge
transformations, eqs.~(\ref{eq:d.6}) and (\ref{eq:d.7}), are written in infinitesimal form as
\begin{eqnarray}
\delta \psi (q,p)&=&-i\lambda \star \psi (q,p)\nonumber  \\
\delta \bar{\psi}(q,p)&=&i\bar{\psi}(p,q)\star \lambda .   \label{eq:d.8}
\end{eqnarray}
Applying these transformations to eq. (\ref{eq:d.5}), leads to the relation
\begin{eqnarray}
\delta \mathcal{L} &=&-\frac{i}{2}\bar{\psi}(p,q)\gamma ^{\mu }\star \lambda \star \left(
p_{\mu }\star \psi (p,q)\right) -\frac{1}{2}\bar{\psi}(p,q)\gamma ^{\mu
}\star \delta \left[ \left( p_{\mu }\star \psi (p,q)\right) \right]  \notag
\\
&&-\frac{1}{2}\delta \left[ \left( \bar{\psi}(p,q)\star p_{\mu }\right) %
\right] \star \gamma ^{\mu }\psi (p,q)+\frac{i}{2}\left( \bar{\psi}%
(p,q)\star p_{\mu }\right) \gamma ^{\mu }\star \lambda \star \psi (p,q).
\label{eq:d.9a}
\end{eqnarray}%
To facilitate visualization of calculations in eq. (\ref{eq:d.9a}), the terms
in the Lagrangian $\delta \left( p^{\mu }\star \psi (p,q)\right) $ and $%
\delta \left[ \left( \bar{\psi}(p,q)\star p_{\mu }\right) \right] $, are
calculated separately. Then the term $\delta \left( p^{\mu }\star \psi
(p,q)\right) $ has the form
\begin{eqnarray}
T_{1} &\equiv &\bar{\psi}(p,q)\gamma ^{\mu }\star \delta \left[ \left(
p_{\mu }\star \psi (p,q)\right) \right]  \notag \\
&=&\bar{\psi}(p,q)\gamma ^{\mu }\star \left( p_{\mu }-\frac{i}{2}\frac{%
\partial }{\partial q^{\mu }}\right) \left( -i\lambda \star \psi (p,q)\right)
\notag \\
&=&\bar{\psi}(p,q)\gamma ^{\mu }\star \left( -ip_{\mu }(\lambda \star \psi
(p,q))+\frac{i^{2}}{2}\frac{\partial \lambda }{\partial q^{\mu }}\star \psi
(p,q)+\frac{i^{2}}{2}\lambda \star \frac{\partial \psi (p,q)}{\partial
q^{\mu }}\right) .  \notag
\end{eqnarray}%
Using identities~\cite{Blaszak}
\begin{eqnarray}
p_{k}(f\star g) &=&f\star (p_{k}g)-\frac{i}{2}(\partial _{qk}f)\star g
\notag \\
&=&(p_{k}f)\star g+\frac{i}{2}f\star (\partial _{qk}g),  \label{eq:d.10}
\end{eqnarray}%
the term $\delta \left( p^{\mu }\star \psi (p,q)\right) $ is given by
\begin{equation}
T_{1}=-i\bar{\psi}(p,q)\gamma ^{\mu }\star \lambda \star \left( p_{\mu
}\star \psi (p,q)\right) -\bar{\psi}(p,q)\gamma ^{\mu }\star \frac{\partial
\lambda }{\partial q^{\mu }}\star \psi (p,q).\label{eq:d.10a}
\end{equation}%
Using the same procedure for the term $\delta \left( \bar{\psi}(p,q)\star p_{\mu
}\right) $ leads to
\begin{eqnarray}
T_{2} &\equiv &\delta \left[ \left( \bar{\psi}(p,q)\star p_{\mu }\right) %
\right] \star \gamma ^{\mu }\psi (p,q)  \notag \\
&=&\left( \bar{\psi}(p,q)\star p_{\mu }\right) \star \lambda \star \gamma
^{\mu }\psi (p,q)i-\bar{\psi}(p,q)\star \frac{\partial \lambda }{\partial
q^{\mu }}\star \gamma ^{\mu }\psi (p,q).  \label{eq:d.10b}
\end{eqnarray}%
Then eq. (\ref{eq:d.9a}) has the form
\begin{equation}
\delta \mathcal{L}=\bar{\psi}(p,q)\star \frac{\partial \lambda }{\partial
q^{\mu }}\star \gamma ^{\mu }\psi (p,q).\label{eq:d.10c}
\end{equation}%
This equation represents a gauge transformation in a general form for the Lagrangian, eq.~(\ref{eq:d.5}). If the gauge transformation is
considered global, where $\lambda $ is a constant, the Lagrangian becomes
invariant, i.e.~( $\delta \mathcal{L}=0$ ). However, if $\lambda $ is a
spacetime function, i.e , $\lambda =\lambda (q,p)$, a local gauge
transformation or gauge transformation of the second kind emerges. An extra
term is generated in the above equation $\frac{\partial \lambda }{\partial
q^{\mu }}\star \psi (p,q)$, making the Lagrangian non-invariant. To restore
gauge invariance a new field, a 4-vector potential $A_{\mu }$ directly
connected with $\psi (q,p)$ is introduced. In this case, eq.~(\ref{eq:d.10a}%
) is rewritten as $\delta \left( p_{\mu }\star \psi (p,q)+\mathcal{L}%
_{a}\right) $, where the extra term $\mathcal{L}_{a}$ is defined as
\begin{equation}
\mathcal{L}_{a}=A_{\mu }\star \psi (p,q).  \label{eq:d.16}
\end{equation}%
The local gauge transformation leads to
\begin{eqnarray}
\delta \left( p_{\mu }\star \psi (p,q)+\mathcal{L}_{a}\right) &=&p_{\mu
}\star \left( \delta \psi (p,q)\right) +\left( \delta \mathcal{L}_{a}\right)
\notag \\
&=&-i\lambda \star \left( p_{\mu }\star \psi (p,q)\right) -\frac{\partial
\lambda }{\partial q^{\mu }}\star \psi (p,q)+\left( \delta A_{\mu }\right)
\star \psi (p,q)  \notag \\
&&-iA_{\mu }\star \lambda \star \psi (p,q).  \label{eq:d.17}
\end{eqnarray}%
This equation has extra terms. Two terms are provided by
Lagrangian $\mathcal{L}_{a}$, where $(\delta A_{\mu })$ is still unknown,
and the other term $\frac{\partial \lambda }{\partial q^{\mu }}$ leads to
the term $\left( p_{\mu }\star \psi (p,q)\right) $. In order to cancel the
extra term, $(\delta A_{\mu })$ takes the form
\begin{eqnarray}
A_{\mu }\rightarrow A_{\mu }^{\prime } &=&A_{\mu }+iA_{\mu }\star \lambda
-i\lambda \star A^{\mu }-\frac{i}{e}\frac{\partial \lambda }{\partial q^{\mu
}}  \notag \\
\delta A_{\mu } &=&iA_{\mu }\star \lambda -i\lambda \star A_{\mu }-\frac{i}{e%
}\frac{\partial \lambda }{\partial q^{\mu }}  \notag \\
&=&i\left\{ A_{\mu },\lambda \right\} _{M}-\frac{i}{e}\frac{\partial \lambda
}{\partial q^{\mu }},  \label{eq:d.18}
\end{eqnarray}%
where $e$ represent the charge of the particle, and $\left\{ a,b\right\}
_{M}=a\star b-b\star a$ is the Moyal bracket. Then defining the operator
\begin{equation}
D_{\mu }\star =p_{\mu }\star +ieA_{\mu }\star ,  \label{eq:d.19}
\end{equation}%
which directly couples to the field $\psi (q,p)$. Applying the gauge
transformation of second kind to $D_{\mu }\star \psi (q,p)$, leads to
\begin{eqnarray*}
\delta \left( D_{\mu }\star \psi (p,q)\right) &=&-i\lambda \star \left( D_{\mu }\star \psi (p,q)\right) .
\end{eqnarray*}%
Then the operator defined in eq.~(\ref{eq:d.19}) obeys the covariant
transformation rule. Similarly in eq. (\ref{eq:d.10b}), there is the
operator $\delta \left( \bar{\psi}(p,q)\star p_{\mu }+\mathcal{L}_{b}\right)
$ where the extra term $\mathcal{L}_{b}$ is defined by 
\begin{equation}
\mathcal{L}_{b}=\bar{\psi}(p,q)\star A_{\mu }.  \label{eq:d.21}
\end{equation}%
A local gauge transformation leads to
\begin{eqnarray}
\delta \left( \bar{\psi}(p,q)\star p_{\mu }+\mathcal{L}_{b}\right) &=&\left( \bar{\psi}(p,q)\star p_{\mu }\right) \star \lambda i-\bar{\psi}%
(p,q)\star \frac{\partial \lambda }{\partial q^{\mu }}  \notag \\
&&+\left( i\bar{\psi}(p,q)\star \lambda \right) \star A_{\mu }+\bar{\psi}%
(p,q)\star \left( \delta A_{\mu }\right) .  \notag
\end{eqnarray}%
Using the definition of $\left( \delta A_{\mu }\right) $ an operator is defined as
\begin{equation}
\star D_{\mu }=\star p_{\mu }+\star ieA_{\mu }.  \label{eq:d.22}
\end{equation}%
%
%
Then the operator, $\bar{\psi}(p,q)\star D_{\mu }$, under a gauge
transformation of the second kind is given by
\begin{eqnarray}
\delta \left( \bar{\psi}(p,q)\star D_{\mu }\right) &=&\left( \bar{\psi}(p,q)\star D_{\mu }\right) \star \lambda i.  \notag
\end{eqnarray}%
This satisfies the rule of a covariant transformation. Rewriting the
Lagrangian, eq.~(\ref{eq:d.5}), in terms of operators $D_{\mu }\star $ and $%
\star D_{\mu }$, leads to
\begin{eqnarray}
\mathcal{L} &=&-\frac{1}{2}\left[ \bar{\psi}(p,q)\gamma ^{\mu }\star D_{\mu
}\star \psi (p,q)+\left( \bar{\psi}(p,q)\star D_{\mu }\right) \star \gamma
^{\mu }\psi (p,q)\right]  \notag \\
&&-m\bar{\psi}(p,q)\star \psi (p,q)+2p_{\mu }\bar{\psi}(p,q)\gamma ^{\mu
}\star \psi (p,q).  \label{eq:d.23}
\end{eqnarray}%
The new Lagrangian under a local gauge transformation, becomes
\begin{eqnarray}
\delta \mathcal{L} &=&-\frac{1}{2}\bigg[\delta (\bar{\psi}(p,q))\gamma ^{\mu }\star (D_{\mu
}\star \psi (p,q))+\bar{\psi}(p,q)\gamma ^{\mu }\star \delta (D_{\mu }\star
\psi (p,q))  \notag \\
&&+\delta \left( \bar{\psi}(p,q)\star D_{\mu }\right) \star \gamma ^{\mu
}\psi (p,q)+\left( \bar{\psi}(p,q)\star D_{\mu }\right) \star \gamma ^{\mu
}\delta (\psi (p,q))\bigg]  =0
\end{eqnarray}%
Therefore the Lagrangian described by eq.~(\ref{eq:d.23}) is invariant under a
local gauge transformation. However, the introduction of the field $A_{\mu }$
gives rise to the electromagnetic interaction. To eliminate this
interaction, another term is introduced in the Lagrangian for the
electromagnetic field. The first step leads to finding the electromagnetic
tensor in phase space. Starting with the relation $[D_{\mu }\star ,D_{\nu
}\star ]$, leads to
\begin{eqnarray}
\lbrack D_{\mu }\star ,D_{\nu }\star ]\psi (q,p) &=&\left[ p_{\mu }\star ,p_{\nu }\star \right] \psi (q,p)+\left[ p_{\mu
}\star ,ieA_{\nu }\star \right] \psi (q,p)  \notag \\
&&+\left[ ieA_{\mu }\star ,p_{\nu }\star \right] \psi (q,p)+\left[ ieA_{\mu
}\star ,ieA_{\nu }\star \right] \psi (q,p).  \label{eq:d.23a}
\end{eqnarray}%
For a better understanding, each term is calculated separately. The first
term is given as
\begin{eqnarray}
T_{3} &=&\left[ p_{\mu }\star ,p_{\nu }\star \right] \psi (q,p)  \notag \\
&=&\left[ p_{\mu }-\frac{i\hbar }{2}\frac{\partial }{\partial q^{\mu }}%
,p_{\nu }-\frac{i\hbar }{2}\frac{\partial }{\partial q^{\nu }}\right] \psi
(q,p)=0.  \notag
\end{eqnarray}%
The second term in eq. (\ref{eq:d.23a}) is
\begin{eqnarray}
T_{4} &=&\left[ p_{\mu }\star ,ieA_{\nu }\star \right] \psi (q,p)  \notag \\
&=&iep_{\mu }A_{\nu }\star \psi (q,p)-ieA_{\nu }\star (p_{\mu }\psi (q,p))-%
\frac{i^{2}e\hbar }{2}\frac{\partial A_{\nu }}{\partial q^{\mu }}\star \psi
(q,p).  \notag
\end{eqnarray}%
Using the identity, eq.(\ref{eq:d.10}), the second term in $T_{4}$ is
%
%
\begin{equation}
A_{\nu }\star (p_{\mu }\psi (q,p))=p_{\mu }\left( A_{\nu }\star \psi
(p,q)\right) +\frac{i\hbar }{2}\frac{\partial A_{\nu }}{\partial q^{\mu }}%
\star \psi (p,q).  \notag
\end{equation}%
Then the $T_{4}$ term becomes
\begin{equation*}
T_{4}=-i^{2}e\hbar \frac{\partial A_{\nu }}{\partial q^{\mu }}\star \psi
(q,p).
\end{equation*}%
By analogy the third term of eq.~(\ref{eq:d.23a}%
) is written as 
\begin{eqnarray}
T_{5}&=&i^{2}e\hbar \frac{\partial A_{\mu }}{\partial q^{\nu }}\star \psi (q,p),
\notag
\end{eqnarray}%
and the last term of eq. (\ref{eq:d.23a}) becomes
\begin{eqnarray}
T_{6} &=&\left[ ieA_{\mu }\star ,ieA_{\nu }\star \right] \psi (q,p)  \notag
\\
&=&i^{2}e^{2}\left\{ A_{\mu },A_{\nu }\right\} _{M}\star \psi (q,p).  \notag
\end{eqnarray}%
Therefore the expression for $[D_{\mu }\star ,D_{\nu }\star ]$ has the form
\begin{equation*}
\lbrack D_{\mu }\star ,D_{\nu }\star ]\psi (q,p)=\hbar e\left( \frac{%
\partial A_{\nu }}{\partial q^{\mu }}-\frac{\partial A_{\mu }}{\partial
q^{\nu }}\right) \star \psi (q,p)+i^{2}e^{2}\left\{ A_{\mu },A_{\nu
}\right\} _{M}\star \psi (q,p).
\end{equation*}%
Then the electromagnetic tensor is defined as
\begin{equation}
\mathcal{F}_{\mu \nu }=\frac{\partial A_{\mu }}{\partial q^{\nu }}-\frac{%
\partial A_{\nu }}{\partial q^{\mu }}+e\left\{ A_{\mu },A_{\nu }\right\}
_{M},  \label{eq:d.23c}
\end{equation}%
where $\mathcal{F}_{\mu \nu }$ is invariant. The invariance of $\mathcal{F}%
_{\mu \nu }$, is demonstrated using the expression
\begin{equation*}
\delta \mathcal{F}_{\mu \nu }=\delta \left( \frac{\partial A_{\mu }}{%
\partial q^{\nu }}-\frac{\partial A_{\nu }}{\partial q^{\mu }}+e\left\{
A_{\mu },A_{\nu }\right\} _{M}\right) .
\end{equation*}%
Now each term is calculated separately. Using eq. (\ref{eq:d.18}) we get
\begin{eqnarray}
\frac{\partial \delta A_{\mu }}{\partial q^{\nu }} &=&i\frac{\partial A_{\mu }}{\partial q^{\nu }}\star \lambda +iA_{\mu }\star
\frac{\partial \lambda }{\partial q^{\nu }}-i\frac{\partial \lambda }{%
\partial q^{\nu }}\star A_{\mu }-i\lambda \star \frac{\partial A_{\mu }}{%
\partial q^{\nu }}-\frac{i}{e}\frac{\partial ^{2}\lambda }{\partial q^{\nu
}\partial q^{\mu }},  \notag
\end{eqnarray}%
and
\begin{eqnarray}
-\frac{\partial \delta A_{\nu }}{\partial q^{\mu }} &=&-i\frac{\partial A_{\nu }}{\partial q^{\mu }}\star \lambda -iA_{\nu
}\star \frac{\partial \lambda }{\partial q^{\mu }}+i\frac{\partial \lambda }{%
\partial q^{\mu }}\star A_{\nu }+i\lambda \star \frac{\partial A_{\nu }}{%
\partial q^{\mu }}+\frac{i}{e}\frac{\partial ^{2}\lambda }{\partial q^{\mu
}\partial q^{\nu }}.  \notag
\end{eqnarray}%
The last term takes the form
\begin{eqnarray}
\delta \left( eA_{\mu }\star A_{\nu }-eA_{\nu }\star A_{\mu }\right)
&=&-ie\lambda \star A_{\mu }\star A_{\nu }-i\frac{\partial \lambda }{%
\partial q^{\mu }}\star A_{\nu }+ieA_{\mu }\star A_{\nu }\star \lambda
-iA_{\mu }\star \frac{\partial \lambda }{\partial q^{\nu }}  \notag \\
&&+ie\lambda \star A_{\nu }\star A_{\mu }+i\frac{\partial \lambda }{\partial
q^{\nu }}\star A_{\mu }-ieA_{\nu }\star A_{\mu }\star \lambda +iA_{\nu
}\star \frac{\partial \lambda }{\partial q^{\mu }}.  \notag
\end{eqnarray}%
%
Then the electromagnetic tensor under local gauge
transformation is given as
\begin{equation*}
\delta \mathcal{F}_{\mu \nu }=-i\lambda \star \left( \mathcal{F}_{\mu \nu
}\right) +\left( \mathcal{F}_{\mu \nu }\right) \star \lambda i.
\end{equation*}%
Using eq.~(\ref{eq:d.4.1}) leads to
\begin{equation}
\int \delta \mathcal{F}_{\mu \nu }=-i\int \lambda \star \mathcal{F}_{\mu \nu
}+i\int \mathcal{F}_{\mu \nu }\star \lambda =0.  \notag
\end{equation}%
%
Defining the last term of eq.~(\ref{eq:d.23}) as
\begin{equation*}
\mathcal{L}_{c}=-\frac{1}{4}\mathcal{F}^{\mu \nu }\mathcal{F}_{\mu \nu },
\end{equation*}%
the final Lagrangian invariant under local gauge transformation has the form
\begin{eqnarray}
\mathcal{L}_{final} &=&-\frac{1}{2}\left[ \bar{\psi}(p,q)\gamma ^{\mu }\star
D_{\mu }\star \psi (p,q)+\left( \bar{\psi}(p,q)\star D_{\mu }\right) \star
\gamma ^{\mu }\psi (p,q)\right]  \notag \\
&-&m\bar{\psi}(p,q)\star \psi (p,q)+2p_{\mu }\bar{\psi}(p,q)\gamma ^{\mu
}\star \psi (p,q)-\frac{1}{4}\mathcal{F}^{\mu \nu }\mathcal{F}_{\mu \nu }.
\label{eq.B22}
\end{eqnarray}%
This demonstrates a mapping similar to that achieved for the Seiberg-Witten
gauge for non-commutative fields.

Our goal now is to study the gauge transformation considering the Moyal-Weyl
star-product in approximate form, i.e. expanding in power series the star
product of general gauge transformation to the zeroth order in $\hbar$. Then
eq.~(\ref{eq:d.5}) is written as
\begin{eqnarray}
S&=&\int \mathcal{L}\, d^4p\, d^4q  \notag \\
&=&\int\bigg[-\frac{1}{2} \left[ \bar{\psi}(p,q)\gamma^{\mu } \left( p_\mu
\star \psi (p,q) \right) + \left(\bar{\psi}(p,q)\star p_\mu \right)
\gamma^{\mu } \psi (p,q) \right]  \notag \\
&&- m \bar{\psi}(p,q) \psi (p,q) + 2 p_{\mu } \bar{\psi}(p,q) \gamma^{\mu }
\psi (p,q)\bigg] d^4p d^4q  \notag \\
&=&\int\bigg[ -\frac{i}{4} \left( \frac{\partial \bar{\psi} (p,q)}{ \partial
q^\mu} \gamma ^{\mu }\psi(q,p) - \bar{\psi} (p,q) \gamma ^{\mu } \frac{%
\partial \psi (p,q)}{ \partial q^\mu} \right)  \notag \\
&&-\bar{\psi} (p,q)\left( m -\gamma ^{\mu }p_{\mu }\right)\psi (p,q) \bigg] %
d^4p d^4q.  \label{eq:di.1}
\end{eqnarray}
Expanding the star product of the gauge transformation in the general form,
eqs. (\ref{eq:d.6}) and (\ref{eq:d.7}), in the zeroth order lead to
\begin{eqnarray}
\psi(q,p) &\rightarrow& e^{-i \lambda} \psi(q,p),  \notag \\
\bar{\psi}(q,p)&\rightarrow& e^{i \lambda} \bar{\psi}(q,p).  \notag
\end{eqnarray}%
Infinitesimal gauge transformation is written as
\begin{align}
\delta \psi(q,p)= -i \lambda\psi(q,p) \\
\delta \bar{\psi}(q,p)= i \lambda\bar{\psi}(q,p).  \label{eq:di.2}
\end{align}
%
Applying this transformation to eq.~(\ref{eq:di.1}) leads
to
\begin{eqnarray}
\delta \mathcal{L}&=&\delta \Bigg\{ -\frac{i}{4} \left( \frac{\partial \bar{%
\psi} (p,q)}{ \partial q^\mu} \gamma ^{\mu }\psi(q,p) - \bar{\psi} (p,q)
\gamma ^{\mu } \frac{\partial \psi (p,q)}{ \partial q^\mu} \right) -\bar{\psi%
} (p,q)\left( m -\gamma ^{\mu }p_{\mu }\right)\psi (p,q) \Bigg\}  \notag \\
&=& \frac{1}{2} \frac{\partial \lambda }{ \partial q^\mu} \bar{\psi} (p,q)
\gamma ^{\mu }\psi(q,p).  \label{eq:di.3}
\end{eqnarray}
Considering the global gauge transformation of eq.~(\ref{eq:di.3})
leads to $\delta \mathcal{L}=0$, i.e. an invariant Lagrangian. The Noether theorem gives a conserved current,
\begin{eqnarray}
J^{\mu}& =& \frac{\partial \mathcal{L}}{\partial\left(\frac{\partial
\psi(q,p) }{\partial q^\mu} \right)} (-i\psi(q,p)) +\frac{\partial \mathcal{L%
}}{\partial\left( \frac{\partial \bar{\psi}(q,p) }{\partial q^\mu} \right)}
(i\bar{\psi}(q,p))  \notag \\
&=& \frac{1}{2} \bar{\psi} (p,q) \gamma ^{\mu } \psi(q,p).  \label{eq:di.4}
\end{eqnarray}
%
For local gauge transformation, eq.~(\ref{eq:di.3}) is written as
\begin{eqnarray}
\delta \mathcal{L} &=& \frac{1}{2} \frac{\partial \lambda }{ \partial q^\mu}
\bar{\psi} (p,q) \gamma ^{\mu }\psi(q,p) = \frac{\partial \lambda }{
\partial q^\mu} J^{\mu }.  \notag
\end{eqnarray}
Therefore the Lagrangian, eq. (\ref{eq:di.1}), is not invariant under local
gauge transformation. To restore local gauge invariance, a new four-vector, $%
A_{\mu}$, coupling directly to the current $J^{\mu}$ is introduced
\begin{eqnarray}
\mathcal{L}&=& -\frac{i}{4} \left( \frac{\partial \bar{\psi} (p,q)}{
\partial q^\mu} \gamma ^{\mu }\psi(q,p) - \bar{\psi} (p,q) \gamma ^{\mu }
\frac{\partial \psi (p,q)}{ \partial q^\mu} \right) -\bar{\psi} (p,q)\left(
m -\gamma ^{\mu }p_{\mu }\right)\psi (p,q) +\mathcal{L}_1,  \label{eq:di.6}
\end{eqnarray}
where
\begin{eqnarray}
\mathcal{L}_1 =- \frac{i}{2} e \bar{\psi} (p,q) \gamma ^{\mu }\psi(q,p)
A_\mu,  \label{eq:d.31}
\end{eqnarray}
and $e$ is the charge of the particle. Using the definition of $%
(\delta A_\mu) $, eq.~(\ref{eq:d.18}), and expanding in power series in $%
\hbar$ leads to
\begin{eqnarray}
A^{\mu} \rightarrow {A^{\prime}}^{\mu}&=& A^{\mu}+ i A^{\mu} \star \lambda -i
\lambda \star A^{\mu} -\frac{i}{e} \frac{\partial \lambda}{\partial q_\mu}
\notag \\
&=& A^{\mu} -\frac{i}{e} \frac{\partial \lambda }{\partial q_\mu}.  \notag
\end{eqnarray}%
Therefore $\delta A_\mu$ has the form
\begin{eqnarray}
\delta A_\mu&=&- \frac{i}{e} \frac{\partial \lambda}{\partial q^\mu}.
\label{eq:d.32}
\end{eqnarray}
 Lagrangian, eq.~(\ref{eq:di.6}), under a local gauge
transformation becomes 
\begin{eqnarray}
\delta \mathcal{L}&=&\delta \Bigg\{ -\frac{i}{4} \left( \frac{\partial \bar{%
\psi} (p,q)}{ \partial q^\mu} \gamma ^{\mu }\psi(q,p) - \bar{\psi} (p,q)
\gamma ^{\mu } \frac{\partial \psi (p,q)}{ \partial q^\mu} \right) -\bar{\psi%
} (p,q)\left( m -\gamma ^{\mu }p_{\mu }\right)\psi (p,q) \Bigg\}  \notag \\
&+& \delta (\mathcal{L}_1)  \notag \\
&=& \frac{1}{2} \bar{\psi} (p,q) \gamma ^{\mu }\psi(q,p) \frac{\partial
\lambda }{ \partial q^\mu} - \delta \left[\frac{i}{2} e \bar{\psi} (p,q)
\gamma ^{\mu }\psi(q,p) A_\mu\right]  \notag \\
&=& 0,
\end{eqnarray}
To demonstrate its invariance, ${\cal L}$ is written
as
\begin{eqnarray}
\mathcal{L}_{t}&=&-\frac{1}{2} \bigg[ \bar{\psi}(p,q)\gamma^{\mu } \left( D_\mu \psi
(p,q)\right) + \left( \bar{D_\mu} \bar{\psi} (p,q) \right) \gamma^{\mu }
\psi (p,q) \bigg] - m \bar{\psi}(p,q) \psi (p,q)  \notag \\
&&+ 2 p_{\mu } \bar{\psi}(p,q) \gamma^{\mu } \psi (p,q),  \label{eq:d.33}
\end{eqnarray}
where the operators are defined as
\begin{equation}
D_\mu \psi(p,q) = \left( p_\mu - \frac{i}{2}\frac{\partial }{ \partial q^\mu}
+ \frac{i}{2} e A_\mu \right) \psi (p,q),  \label{eq:d.34}
\end{equation}
and
\begin{equation}
\bar{D}_\mu \bar{\psi}(p,q) = \left( p_\mu + \frac{i}{2}\frac{\partial }{
\partial q^\mu} + \frac{i}{2}eA_\mu \right) \bar{\psi} (p,q).
\label{eq:d.35}
\end{equation}
Under local gauge transformation, the operator $D_\mu $ becomes covariant
as
\begin{eqnarray*}
\delta(D_\mu \psi(p,q)) &=&-i \lambda D_\mu \psi (p,q).  \label{eq:d.36}
\end{eqnarray*}
In order to eliminate the electromagnetic interaction due to the introduction of $%
A_\mu$ power series expansion to zero order in $\hbar$, the star product of
eq.~(\ref{eq:d.23c}), is
\begin{equation}
F_{\mu\nu} = \frac{\partial A_{\nu}}{\partial q^{\mu}} -\frac{\partial
A_{\mu}}{\partial q^{\nu}}.  \label{eq:d.37}
\end{equation}
The invariance of $F_{\mu\nu}$ is
\begin{eqnarray}
\delta F_{\mu\nu} &=& \frac{\partial (\delta A_{\nu})}{\partial q^{\mu}} -%
\frac{\partial (\delta A_{\mu})}{\partial q^{\nu}} =0,  \notag
\end{eqnarray}
where eq. (\ref{eq:d.32}) is used. Therefore, the last term of eq.~(\ref{eq:d.33}) leads to
\begin{eqnarray}
\mathcal{L}_{2} = -\frac{1}{4} F^{\mu\nu}F_{\mu\nu},  \label{eq:d.38}
\end{eqnarray}
and the final Lagrangian is 
\begin{eqnarray}
\mathcal{L}_{final} &=& \mathcal{L}_{t} + \mathcal{L}_{2}  \notag \\
&=&-\frac{1}{2} \bigg[ \bar{\psi}(p,q)\gamma^{\mu } \left( D_\mu \psi
(p,q)\right) + \left( \bar{D_\mu} \bar{\psi} (p,q) \right) \gamma^{\mu }
\psi (p,q) \bigg]  \notag \\
&-& m \bar{\psi}(p,q) \psi (p,q) + 2 p_{\mu } \bar{\psi}(p,q) \gamma^{\mu }
\psi (p,q) -\frac{1}{4} F^{\mu\nu}F_{\mu\nu}.  \label{eq:d.39}
\end{eqnarray}
The results show a Lagrangian, defined in phase space, is similar to that obtained in quantum field theory.


\begin{thebibliography}{999}

\bibitem{Wigner} E. Wigner, Phys. Rev. \textbf{40}, 749 (1932).

\bibitem{Heisenberg} W. Pauli, Scientific correspondence, Vol II, p.15, Ed. K.
von Meyenn (Spring-Verlag, Berlin, 1985).

\bibitem{Snyder} H. S. Snyder, Phys. Rev. {\bf 71}, 38 (1947).

\bibitem{Moyal} J. E.  Moyal, Proc. Cambridge Phil. Soc. \textbf{45}, 99 (1949). 

\bibitem{Hillery} M. Hillery, R. F. O
Connel, M. O. Scully and E. P. Wigner, Phys. Rep. \textbf{106}, 121 (1984).




\bibitem{zabo1} R. J. Szabo, Phys. Rep. \textbf{378}, 207 (2003)
\bibitem{Grosse} H. Grosse and R. Wilkenhaar, Comumm.
Math. Phys. {\bf 256}, 305 (2005).
\bibitem{SeiberWitten} N. Seiberg and E. Witten, JHEP \textbf{9909} (1999) [hep-th/9908142].

\bibitem{Minwalla} S. Minwalla, M. Van Raamsdonk and N. Seiberg, J. High
Energy Phys. {\bf 02}, 020 (2000).


\bibitem{Langmann} E. Langmann and R. J. Szabo, Phys. Lett. B {\bf 533}, 168 (2002).

\bibitem{Magnen} J. Magnen, V. Rivasseau and A. Tanasa, Eur. Phys. Lett. {\bf 86},
11001 (2009).


\bibitem{Mariz} T. Mariz, J. R. Nascimento and V. O. Rivellis, Phys. Rev. D {\bf 75}, 025020 (2007).

\bibitem{Costa} M. L. Costa, A. R. Queiroz and A. E. Santana, Int. J. Mod. Phys. A {\bf 25}, 3209 (2010).

\bibitem{Gurau} R. Gurau, A. P. C. Malbouisson, V. Rivasseau and A.Tanasa,  Lett. Math. Phys. {\bf 81}, 161 (2007).

\bibitem{Kalau} W. Kalau and M. Walze, J. Geom. Phys. {\bf 16}, 327 (1955).

\bibitem{Kastler} D. Kastler, Commun. Math. Phys. {\bf 166}, 633 (1995).

\bibitem{Girotti} H. O. Girotti, M. Gomes, V. O. Rivelles and A. J. da
Silva,  Nucl.Phys. B {\bf 587}, 299 (2000).

\bibitem{Belissard} J. Belissard, A. van Elst and H. Schulz-Baldes, J. Math.
Phys. {\bf 35}, 53 (1994).

\bibitem{Yu} H. Yu and  Bo-Qiang Ma, Mod. Phys. Lett. A \textbf{32}, 1750030 (2017).

\bibitem{Hammil} B. Hamil, Mod. Phys. Lett. A \textbf{33}, 1850017 (2018).

\bibitem{Amorim0} R. G. G. Amorim, F. C. Khanna, A. P. C. Malbouisson, J.
M. C. Malbouisson and A. E. Santana, Int. J. Mod. Phys.A \textbf{30}, 1550135
(2015).

\bibitem{Amorim19} R. G. G. Amorim, F. C. Khanna, A. P. C. Malbouisson, J.
M. C. Malbouisson and A. E. Santana, Int. J. Mod. Phys.A \textbf{34}, 1950037 
(2019).

\bibitem{Oliveira} M. D. Oliveira, M. C. B. Fernandes, F. C. Khanna, A. E.
Santana and J. D. M.Vianna Ann. Phy. \textbf{312}, 492 (2004). %

\bibitem{Ronni} R. G. G. Amorim , M. C. B. Fernandes , F. C. Khanna, A. E.
Santana and J. D. M. Vianna, Phy. Lett. A \textbf{361}, 464 (2007). %


\bibitem{wig3} Y.S. Kim, M.E. Noz, \textit{Phase Space Picture and Quantum
Mechanics - Group Theoretical Approach }(W. Scientific, London, 1991).

\bibitem{wig4} T. Curtright, D. Fairlie, C. Zachos, Phys. Rev. D \textbf{58}%
, 25002 (1998).

\bibitem{2gal1} D. Galetti and A.F.R. de Toledo Piza, Physica A \textbf{214}%
, 207 (1995).

\bibitem{2davido1} L.G. Lutterbach and L. Davidovich, Phys. Rev. Lett.
\textbf{78}, 2547 (1997).

\bibitem{2kha1} A.E. Santana, A. Matos Neto, J.D.M. Vianna and F.C. Khanna,
Physica A \textbf{280}, 405 (2000).



\bibitem{Ronni001} R. G. G. Amorim, S. C. Ulhoa and E. O. Silva, Braz. J. Phys. \textbf{45}, 664 (2015). %

\bibitem{Ronni002} R. G. G. Amorim , M. C. B. Fernandes , F. C. Khanna , A. E.
Santana and J. D. M. Vianna, Int. J. Mod. Phys. A, \textbf{28}, 1350013 (2013).

\bibitem{Cruz} J. S. da Cruz Filho, R. G. G. Amorim, S. C. Ulhoa, F. C.
Khanna, A. E. Santana and J. D. M. Vianna, Int. J. Mod. Phys. A, \textbf{31},
1650046 (2016).

\bibitem{diracphasespace} R. G. G. Amorim, S. C. Ulhoa and E. O. Silva, Braz. J. Phys. {\bf 45}, 664 (2015).

\bibitem{Blaszak}  M. Blaszak and Z. Domanky, Ann. Phys. \textbf{327}, 167 (2012).

\end{thebibliography}
\end{document}